\documentclass[iop]{emulateapj} 
\usepackage{pifont} 
\usepackage{subfigure}
\shorttitle{CIBER Low Resolution Spectrometer}
\shortauthors{Tsumura et al. (CIBER collaboration)}
\newcommand{\myemail}{{\it tsumura@ir.isas.jaxa.jp}}
\newcommand{\e}{e$^{-}$}
\newcommand{\eps}{e$^{-}/$s}
\newcommand{\nw}{nW/m$^2$/sr}

\begin{document}
\title{The Cosmic Infrared Background Experiment (CIBER): The Low Resolution Spectrometer}
\author{K. Tsumura\altaffilmark{1}, T. Arai\altaffilmark{1,2}, J. Battle\altaffilmark{3}, J. Bock\altaffilmark{3,4}, S. Brown\altaffilmark{5}, A. Cooray\altaffilmark{6}, V. Hristov\altaffilmark{4}, B. Keating\altaffilmark{7}, M. G. Kim\altaffilmark{8}, \\ D. H. Lee\altaffilmark{9}, L. R. Levenson\altaffilmark{4}, K. Lykke\altaffilmark{5}, P. Mason\altaffilmark{4}, T. Matsumoto\altaffilmark{1,8}, S. Matsuura\altaffilmark{1}, K. Murata\altaffilmark{1,10}, U. W. Nam\altaffilmark{9}, T. Renbarger\altaffilmark{7}, A. Smith\altaffilmark{5}, I. Sullivan\altaffilmark{11}, K. Suzuki\altaffilmark{12}, T. Wada\altaffilmark{1}, and M. Zemcov\altaffilmark{3,4}}
\affil{\myemail}
\slugcomment{Accepted by the Astrophysical Journal Supplement}
        
\altaffiltext{1}{Department of Infrared Astrophysics, Institute of Space and Astronoutical Science (ISAS), Japan Aerospace Exploration Agency (JAXA), Sagamihara, Kanagawa 252-5210, Japan}
\altaffiltext{2}{Department of Physics, Graduate School of Science, The University of Tokyo, Tokyo 113-0033, Japan}
\altaffiltext{3}{Jet Propulsion Laboratory (JPL), National Aeronautics and Space Administration (NASA), Pasadena, CA 91109, USA}
\altaffiltext{4}{Department of Astronomy, California Institute of Technology, Pasadena, CA 91125, USA}
\altaffiltext{5}{Optical Technology Division, National Institute of Standards and Technology (NIST), Gaithersburg, MD 20899, USA}
\altaffiltext{6}{Center for Cosmology, University of California, Irvine, Irvine, CA 92697, USA}
\altaffiltext{7}{Department of Physics, University of California, San Diego, San Diego, CA 92093, USA}
\altaffiltext{8}{Department of Physics and Astronomy, Seoul National University, Seoul 151-742, Korea}
\altaffiltext{9}{Korea Astronomy and Space Science Institute (KASI), Daejeon 305-348, Korea}
\altaffiltext{10}{Department of Space and Astronautical Science, School of Physical Sciences, The Graduate University for Advanced Studies, Sagamihara, Kanagawa 252-5210, Japan}
\altaffiltext{11}{Department of Physics, The University of Washington, Seattle, WA 98195, USA}
\altaffiltext{12}{Instrument Development Group of Technical Center, Nagoya University, Nagoya, Aichi 464-8602, Japan}

\begin{abstract}
Absolute spectrophotometric measurements of diffuse radiation at 1 $\mu$m
to 2 $\mu$m are crucial to our understanding of the radiative content of
the Universe from nucleosynthesis since the epoch of reionization, the
composition and structure of the Zodiacal dust cloud in our solar
system, and the diffuse galactic light arising from starlight
scattered by interstellar dust.  The Low Resolution Spectrometer (LRS)
on the rocket-borne Cosmic Infrared Background Experiment (CIBER) is a
$\lambda / \Delta \lambda \sim 15-30$ absolute spectrophotometer
designed to make precision measurements of the absolute near-infrared
sky brightness between 0.75 $\mu$m $< \lambda <$ 2.1 $\mu$m.  This paper
presents the optical, mechanical and electronic design of the LRS, as
well as the ground testing, characterization and calibration
measurements undertaken before flight to verify its performance.  The
LRS is shown to work to specifications, achieving the necessary
optical and sensitivity performance.  We describe our understanding
and control of sources of systematic error for absolute photometry of
the near-infrared extragalactic background light.

\end{abstract}
\keywords{cosmic background radiation --- infrared: diffuse background --- instrumentation: spectrograph --- methods: laboratory --- space vehicles: instruments --- techniques: spectroscopic}

\section{Introduction}
\label{S:intro}

A measurement of the extragalactic background light (EBL) determines the integrated emission of all photon sources since the early universe.  
At near-infrared (NIR) wavelengths, the dominant physical process thought to be responsible for the generation of photons is nucleosynthesis in stars
and gravitational energy released in active galactic nuclei (AGN).
As most stars reside in galaxies, the brightness of the extragalactic
cosmic NIR background (CNIRB) may be constrained by
integrating the light resolved into discrete sources along a line of sight
through the cosmos (e.g.~\citet{Mad00, Tot01, Faz04, Keenan10}).
However, since unresolved or faint sources
of emission may also contribute to the CNIRB, such source counts are
necessarily lower limits to the total emission.  The most complete way
to determine the total CNIRB brightness is direct absolute photometry 
measurements  using suitably designed instruments, 
and the difference between absolute photometry measurements and source counts could reveal a diffuse background from the epoch of reionization.  
Because atmospheric emission at NIR is 100 times more than the total sky brightness, space-borne observations are required for absolute photometry.

Measurements from the Diffuse Infrared Background Explorer (DIRBE) on
the Cosmic Background Explorer ({\sc COBE}) \citep{Hauser98, Cam01} 
and the Near-Infrared Spectrometer (NIRS) on the Infrared
Telescope in Space ({\sc IRTS}) \citep{Matsumoto05} indicate that the
total CNIRB brightness significantly exceeds the brightness determined
from deep galaxy number counts \citep{Mad00, Tot01, Faz04, Keenan10}
after subtraction of the zodiacal light (ZL) from sunlight scattered by interplanetary dust grains.
For example, the best galaxy counts give $14.7\pm 2.4$ \nw\ at J-band ($1.25 \, \mu$m) \citep{Keenan10}
while DIRBE measured $54.0\pm 16.8$ \nw\ at J-band \citep{Cam01}
and IRTS measured $70.1\pm 13.2$ \nw\ at $1.43 \, \mu$m \citep{Matsumoto05} with the DIRBE ZL model \citep{Kelsall98}.
The CNIRB derived from absolute photometry measurements depend critically on the choice of ZL model, 
and the ``strong no-Zodi'' foreground dust model \citep{Wright98} produces significantly lower EBL results, $21\pm 15$ \nw\ at $1.25 \, \mu$m \citep{Wright2001, Levenson07}.
An excess was also measured in optical bands with a combined Hubble Space Telescope ({\sc HST}) and ground-based measurement \citep{Bernstein02, Bernstein05}. 
However, \citet{Mattila03} pointed out that their resulting EBL values should be corrected upwards and, because of large systematic errors, should be understood as upper limits only. 
This was confirmed  by \citet{Bernstein07} whose reanalysis gave a formal new EBL value of $57\pm 33$ \nw\ at 0.8 $\mu$m.
The cause of this discrepancy is unclear; possibilities range from the prosaic, for example residual ZL \citep{Dwe05}, to the profound, such as
Lyman-$\alpha $ emission from the first stars \citep{San02, Sal03, Cooray04, Madau05, Dwe05, Fernandez06}.

On the other hand, a recent re-analysis of Pioneer 10/11 data from outside of the zodiacal cloud gives $7.9\pm 4.0$ \nw\ at 0.44 $\mu$m and $7.7\pm 5.8$ \nw\ at 0.64 $\mu$m \citep{Matsuoka11}, 
and a recent result from the dark cloud shadow method gives $7.2^{+4}_{-2}$ \nw\ at 0.40 $\mu$m and $< 12$ \nw\ at 0.52 $\mu$m \citep{Mattila11}, 
which are more consistent with the source counts at optical wavelengths.
In addition, indirect measurements of the CNIRB from TeV-energy $\gamma$-ray attenuation via pair production are consistent with the source counts \citep{Dwek05, Aha06, Aha07, Mazin07, Raue09},
and these results dispute the theory that a significant fraction of CNIRB comes from EBL. 
In these analysis, the authors calculated attenuation spectra of the blazars in various CNIRB levels by assuming their intrinsic spectra as $dN/dE \propto E^{-\Gamma }$, and compared with observed spectra. 
The general result is that a CNIRB of $>20$ \nw\ requires  $\Gamma <1.5$, which is not consistent with estimates of intrinsic blazar spectra either from theory or from observations of nearby blazars \citep{Aha06}.
These are all viable methods of constraining the brightness of the CNIRB and thus the amplitude of the CNIRB and its origin are controversial.
However, there is no substitute for direct photometry if the instrumental and astrophysical systematic effects can be mitigated with improved knowledge of local foregrounds.
Spectroscopy at wavelengths 1 $\mu$m to 2 $\mu$m is especially important because the CNIRB from reionization is predicted to peak near 1 $\mu$m and fall in brightness at shorter wavelengths.
It is clear that our knowledge of the NIR EBL must improve in order to better constrain the Universe's total emission and thereby models of structure formation and galaxy evolution.

In addition to the EBL science, low resolution absolute spectroscopic
measurements of the background from above the Earth's atmosphere allow
determinations of the ZL and the
diffuse galactic light (DGL) from starlight scattered by interstellar dust grains, neither of which are
well measured near $1 \, \mu$m.  Both of these sources of emission are
truly diffuse in nature and are astrophysically interesting in
their own right.  For example, since the Zodiacal cloud is the
nearest analog to extrasolar dust clouds and debris disks, 
understanding the local Zodiacal dust
environment allows inferences to be made about the nature of distant systems.

The Low Resolution Spectrometer (LRS) on the Cosmic Infrared
Background ExpeRiment (CIBER) \citep{Bock06} is specifically designed to perform
absolute spectrophotometric measurement at NIR wavelength (0.75 $\mu$m $< \lambda <$ 2.1 $\mu$m).  
The LRS has a field of view (FoV) $5.5^{\circ}$ along a slit and
spectral resolution $\lambda / \Delta \lambda \sim$15-30.  The LRS has
successfully flown on CIBER twice, and has yielded good results from
both flights: for example, the LRS previously detected an
unknown silicate feature in the ZL \citep{Tsumura10}.  
Two additional flights of essentially the same configuration of LRS are planned.
In this paper we present the LRS instrument in detail, concentrating on
its optical, mechanical and electrical characteristics\footnote{All uncertainties in the paper are standard uncertainties unless otherwise noted.}.  
In Section \ref{S:instrument} we review the physical properties of the LRS, and
in Section \ref{S:lab} the laboratory testing and characterization of
the LRS is presented.  Finally, we summarize the LRS instrument
and its characterization in Section \ref{S:summary}.

\section{Instrument}
\label{S:instrument}

\subsection{The Cosmic Infrared Background Experiment}
\label{sS:CIBER}

\begin{figure*}
\begin{center}
\includegraphics[scale=0.6,angle=90]{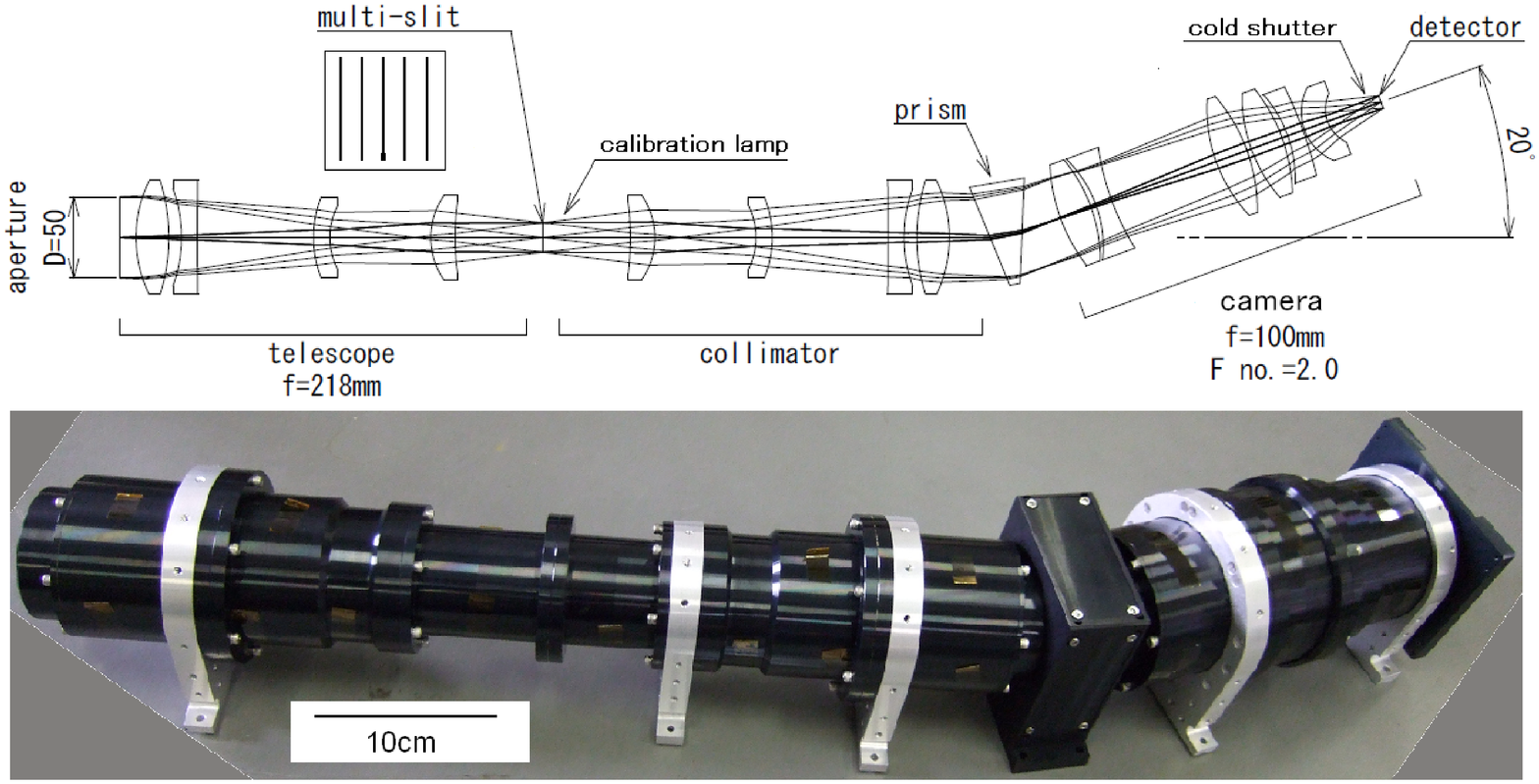}
\end{center}
\caption{Optical schematic and photograph of the LRS.
Light is focused on the multi-slit mask by the telescope, re-expanded by the collimator, dispersed by the prism, and imaged by the camera onto the detector.
The calibration lamp is inserted after the slit, and the cold shutter is attached in front of the detector.  
\label{LRS}}
\vskip 0.5cm
\end{figure*}

CIBER is a rocket-borne instrument designed to search for
fluctuations in the NIR extragalactic background from the first
galaxies, perform direct spectrophotometric measurements of the CNIRB,
and measure the absolute brightness of the ZL cloud using a Fraunhofer line measurement. 
Details of the overall science and instrument package can be found in \citet{Zemcov11}.  
CIBER comprises four optical instruments to achieve its science goals: two
wide-field Imagers \citep{Bock11}, a Narrow Band Spectrometer (NBS; \citet{Renberger11}), 
and the LRS.  These instruments are mounted to a
common optical bench which is cryogenically cooled using an onboard
reservoir of liquid nitrogen to reduce in-band thermal emission below
the detector sensitivity.  For CIBER's first flight, a Terrier-Black
Brant IX rocket \citep{Krause05} carrying the CIBER payload
successfully launched from White Sands Missile Range (WSMR) on 2009 February
25 achieving an apogee of $\sim$330 km and providing 420
seconds of astronomical data.  CIBER's second flight took place at WSMR on 2010
July 10, incorporating several improvements to the instruments, 
and also provided 420 seconds of astronomical data.
The instrument package was successfully recovered for future flights in both cases,
and two additional flights of essentially the same configuration are planned.

Here we briefly review the other two CIBER instruments, which observe simultaneously with the LRS.
The two-color wide-field Imagers are designed to measure fluctuations in the CNIRB.  Both telescopes have a $2^{\circ }\times 2^{\circ }$ FoV 
which allows measurement of the spatial power spectrum on scales at and beyond the predicted reionization peak at 10 arcminutes \citep{Cooray04}. 
The imagers operate in two bands at $1.0 \, \mu$m and $1.6 \, \mu$m so that the reionization signal can be spectrally distinguished from local foregrounds.
The imager pixels are $7\times 7$ square arcseconds so that galaxies can be removed to a sufficient depth to reduce the foreground signal from galaxy clustering \citep{Bock11}.
The NBS is designed as a tipped filter spectrometer \citep{Eather1969} which will measure the absolute intensity of a Fraunhofer line in the ZL to $\sim 1 \,$\% uncertainty.  
The NBS will allow a direct measurement of the ZL brightness in each CIBER field which are needed to determine EBL in absolute photometry measurements \citep{Renberger11}.

\subsection{Low Resolution Spectrometer}
\label{ssS:LRS}
\subsubsection{Optical design}
\label{sS:optical design}

\begin{table}[b]
\caption{Specifications of the Low Resolution Spectrometer\label{LRSspec}}
\begin{center}
\vskip -0.5cm
\begin{tabular}{ll}
\tableline\tableline\noalign{\smallskip}
Characteristic & Value \\
\noalign{\smallskip}\tableline\noalign{\smallskip}
Optics & 14 lenses, 2 filters, 1 prism, 5 slits \\
Aperture & 50 mm \\ 
F number & 2\\
Slit FoV &  5.35 degrees $\times $ 2.8 arcmin \\
Pixel size & 1.36 arcmin $\times $ 1.36 arcmin \\ 
PSF FWHM & $<$ 1.36 arcmin \\
Wavelength range & 0.75$\, \mu$m to 2.1$\, \mu$m \\
Wavelength resolution & $\lambda /\Delta \lambda $=15-30 \\ 
Optical total efficiency & $>$ 0.6 \\
Detector & 256$\times $256 substrate-removed PICNIC \\ 
Detector QE & 0.9 \\  
Median dark current & $<$ 0.6 $e^{-}$/s \\
Median read noise & $<$ 26 $e^{-}$ \\
\noalign{\smallskip}\tableline\noalign{\smallskip}
\end{tabular}
\end{center}
\end{table}

The LRS is designed to obtain the absolute spectrum of the astrophysical
sky brightness for 0.75 $\mu$m $< \lambda <$ 2.1 $\mu$m.  As shown in
Table~\ref{LRSspec}, the LRS is a refractive imaging
spectrometer with a 5-cm aperture designed and fabricated by the Genesia
corporation\footnote{Mitaka Sangyo-Plaza 601, 3-38-4, Shimorenjaku,
  Mitaka, Tokyo, 181-0013, Japan} in Japan.  The most demanding
requirements on the LRS design are driven by the nature of absolute
photometric measurement of faint, diffuse radiation over a wide
wavelength range. The sensitivity and wavelength requirements
lead to a multi-slit spectrometer with a large
FoV ($5.5^{\circ}$) and a large pixel pitch ($40 \,
\mu$m) to maximize the throughput of the system.  
To maximize the
number of independent pixels available to measure surface brightness,
the size of the point spread function (PSF) was designed to be
$< 1$ pixel over the whole array. This design was verified by focus testing (for a discussion
of the latter see section \ref{ssS:focus}).  Furthermore, the LRS is
designed to be cryogenically cycled to $\lesssim 100 \,$K many times
and tolerate both the launch acceleration and vibration, and the space environment experienced by
sounding rocket payloads.  These requirements lead to a simple,
rugged, but precisely optimized mechanical design.  Additionally, the
LRS instrument includes a cold shutter assembly to monitor the
detector dark current for absolute spectroscopy, and a calibration
lamp to confirm the stability of the system during the flight.  
These components are basically common for all four of the CIBER instruments,
and a light emitting diode L10660 peaking at $1450 \,$nm by
Hamamatsu photonics\footnote{http://jp.hamamatsu.com/en/index.html} 
is chosen as the calibration lamp for the LRS to closely match the mean wavelength.
Details of their design and implementation can be found in \citet{Zemcov11}. 

\begin{figure}[h]
\begin{center}
\includegraphics[scale=0.4, angle=90]{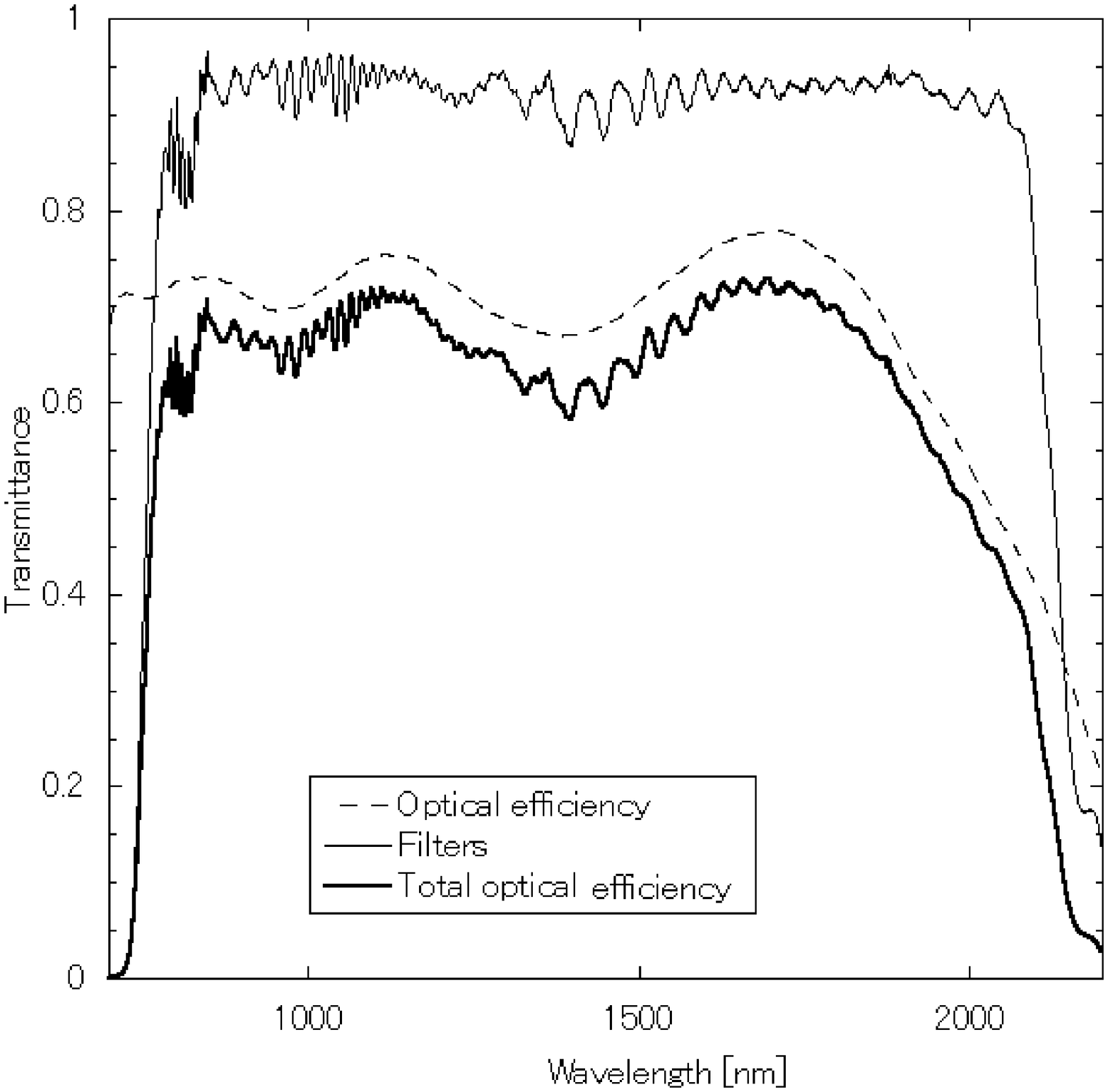}
\end{center}
\caption{The thin solid line shows the combined transmittance of the long and short pass filters, 
the dashed line shows the transmittance of LRS optics including light loss by reflection and absorption by lenses, 
and the thick solid line shows the total efficiency of the LRS optics. \label{filter}}
\end{figure}

The design of the LRS optical elements is based on cryogenic
measurements of the refractive indices of optical materials from \citet{Yamamuro06}.
As shown in Figure~\ref{LRS}, the LRS optics form an initial focus at
a field stop, where a mask with five equally spaced slits is placed.
This slit width is $140\pm 2 \,\mu$m, equivalent to two pixels on the
detector.  The central slit has a small notch whose size on the
detector is 6 pixels by 15 pixels located at the bottom center to
facilitate laboratory testing.  The optics then relay this field stop
back to a collimated beam where a prism disperses incident light
perpendicular to the slit mask direction with a spectral resolution of
$15 \lesssim \lambda /\Delta \lambda \lesssim 30$ depending on wavelength (see section \ref{ssS:pixeltolambda}).  
Finally, this parallel beam is refocused on a $256 \times 256$ substrate-removed HgCdTe PICNIC detector array 
fabricated by Teledyne Technologies Inc.\footnote{http://www.teledyne-si.com/}.
The slit mask is imaged to five separate
2.8 arcminute $\times $ 5.5 degree strips on the sky at the focal
plane.  The edges of the array are not illuminated by the optics and
can be used as a monitor of diffuse stray light falling on the
detector.  The spectral response of the LRS is restricted to 0.75 $\mu$m $< \lambda <$ 2.1 $\mu$m 
by two blocking filters in front of the optics (the effective spectral range is shown in Figure~\ref{filter}). 
In order to detect the peak of the CNIRB, the sensitivity goal of the LRS is 10 \nw\ in a 25 s integration time.

\subsubsection{Electronics}
\label{ssS:electronics}

The LRS focal plane array electronics chain is the same as that of the other three CIBER arrays described in \citet{Zemcov11}.
Here we simply note that the CIBER electronics, which have a common architecture for all four instruments, 
generate the clocking signals for the PICNIC multiplexer, perform the array readout and digitization, 
and are responsible for various housekeeping tasks.  
The digitized array readouts are passed from the CIBER electronics to the NASA-provided payload telemetry module for transmission to ground stations.

An important feature of the infrared arrays is the ability to perform a non-destructive read-out
in which the charge on the detector photo-diode junction is not altered by sampling its value.  
The focal plane infrared arrays on CIBER optics are controlled by a sequence of reset and read-out pulses 
using the method presented in \citet{Hodapp96}.  
The PICNIC array on LRS is separated into four quadrants, and all four quadrants are operated in the same method independently.  
The array is read out at $\sim$4 Hz and then a reset signal is applied after some number of frames \citep{Lee10}.  
Since electrical carriers generated by photons are integrated until the reset,  
an astrophysical image is obtained by calculating the best fit slope to each pixel within a reset interval (``sampling-up-the-ramp'' technique, \citep{Garnett93}).

\section{Laboratory Characterization of the LRS}
\label{S:lab}
\subsection{Detector Performance}
\label{sS:detectorperformance}
\subsubsection{Dark Noise}
\label{ssS:darknoise}

Thermally activated carriers give rise to a continuous stream of electrons in the absence
of light and generate a dark current in HgCdTe detectors.
Because of this, dark current subtraction is
essential for absolute photometry.  In flight, the LRS dark current was 
monitored by taking dark frames with the cold shutter. The temperature
of the CIBER focal plane assemblies was controlled to within $\pm 10
\mu$K/s stability to prevent dark current caused by thermal drift (see \citealt{Zemcov11} for more detail).

The detector noise performance is evaluated by a laboratory dark test.  
In this test, we evaluate several dark frames taken while the cold shutter in front of the detector is closed.
In this shutter closed configuration, any light from the outside does not reach at the array, 
so the best fit slope to each pixel within an integration interval corresponds to the dark current, and the read-out noise can be evaluated from the dispersion from the best-fit line.
In Table \ref{dark}, the average values and their variations of the dark current and read-out noise over the each quadrant or the whole array are shown as representative values.
The resultant dark current and read-out noise of each pixel are consistent with the estimated performance of the detector design (for more detail, see \citealt{Lee10}).

\begin{table}[h]
\caption{Dark current and read noise of the LRS detector\label{dark}}
\begin{center}
\vskip -0.5cm
\begin{tabular}{ccc}
\tableline\tableline\noalign{\smallskip}
\nodata & Dark current (\eps) & Read Noise (\e) \\
\noalign{\smallskip}\tableline\noalign{\smallskip}
Whole array & 0.33 $\pm$ 0.05  & 25.2 $\pm$ 0.5 \\
Quadrant 1 & 0.54 $\pm$ 0.05  & 25.2 $\pm$ 0.5 \\
Quadrant 2 & 0.24 $\pm$ 0.05 & 25.2 $\pm$ 0.5 \\
Quadrant 3 & 0.20 $\pm$ 0.05 & 25.3 $\pm$ 0.5 \\
Quadrant 4 & 0.34 $\pm$ 0.05 & 25.3 $\pm$ 0.5 \\
\noalign{\smallskip}\tableline\noalign{\smallskip}
\end{tabular}
\end{center}
\vskip -0.4cm
\end{table}

\subsubsection{Linearity and Saturation}
\label{ssS:linearity}

\begin{figure*}
\subfigure{
	 \includegraphics[scale=0.4, angle=90]{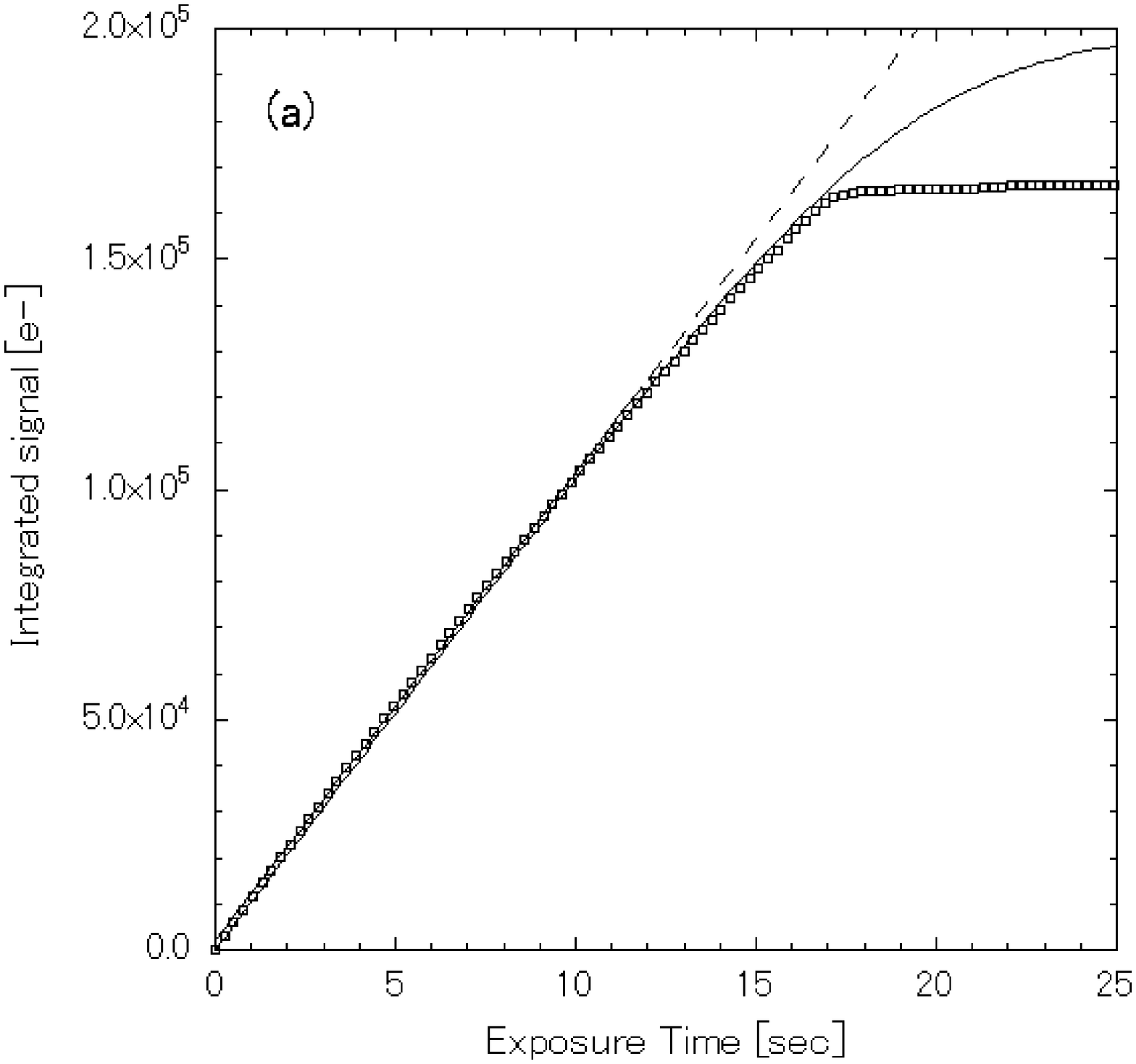}}
\subfigure{
	 \includegraphics[scale=0.4, angle=90]{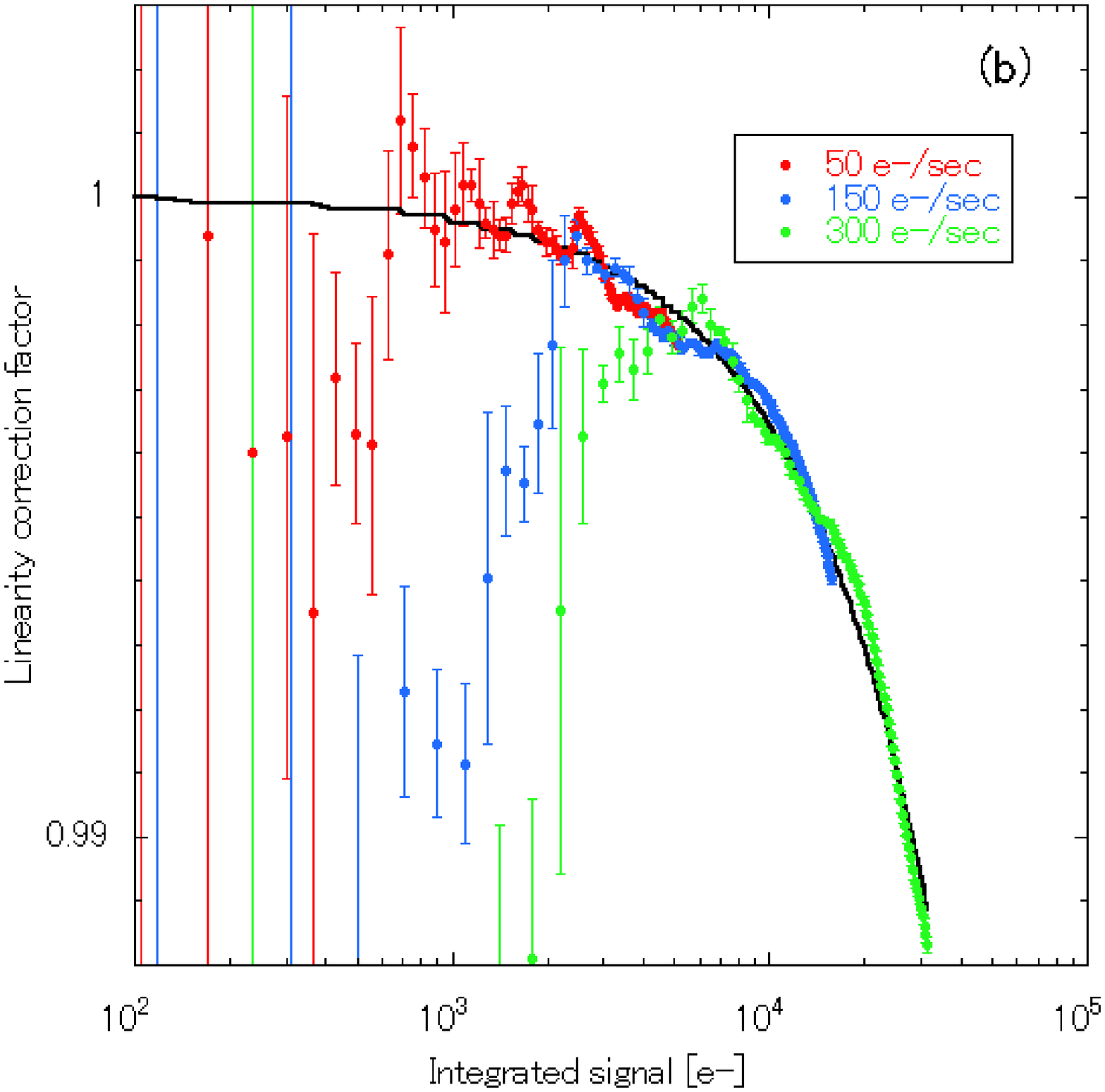}}
\caption{(a) The typical behavior of a detector pixel under saturating illumination.   
The results of the linear fit (broken line) and the fit to the model of \citet{Biesiadzinski11} (solid line) are also shown.   
(b) The linearity correction factor as a function of the integrated signal normalized at 100 \e.
The best fit curve for the correction is also plotted.
The saturation level of $1.65\times 10^5$ \e\ is out of range in horizontal axis.
A  correction up to 1.5\% is required to compare the LRS calibration obtained under laboratory backgrounds to the low-background flight measurements.
\label{saturation}}
\vskip 0.4cm
\end{figure*}

We carried out a photometric test to check linearity and to determine the
saturation properties of the detector array.  In this test, the detector
is illuminated by thermal emission from the laboratory.  
The resulting image has high dynamic range;
pixels at longer wavelengths ($> 1.8 \mu$m) are saturated from thermal radiation,
whereas pixels at the edges of the array are not illuminated by the optics and essentially give dark current.
The first three or four samples following a reset do not follow a linear model because of the
reset anomaly, which gives a large intensity ramp at the leading edge of each quadrant
where the readout electronics is located. Therefore the linear fit excludes the first four samples.
Figure~\ref{saturation} (a) shows an example of the behavior
of a typical pixel with a saturated photo current.
\citet{Biesiadzinski11} introduce a functional form to describe non-linearity, 
the resulting fit is also shown in Figure~\ref{saturation} (a).  
The typical maximum electron capacity is $1.65\times 10^5 \,$\e\ and pixel values deviate 
from the linearity from around $10^5 \,$\e\ over the array.  
For example, for a typical $50 \,$s flight integration a
source brighter than $2000 \,$\eps\ (corresponding to 7 mag at 1 $\mu$m) would begin to cause
non-linearities in the detector.  This threshold is significantly
larger than the estimated LRS diffuse sky brightness signal of $\lesssim 30\,$\eps.
The brightness of sources which induce currents larger than $2000 \,$\eps\ can be derived by
fitting to subsets where the signal depth is $< 10^5 \,$\e.  For
example, the brightest star detected with the LRS during the first
CIBER flight (42Dra, a mag$_J$=2.9, K1.5III star) has a photocurrent of $3\times 10^4 \,$\eps; its
spectrum was derived using data from only the first three seconds of
integration \citep{Tsumura10}.

Though the detector well begins to saturate $\sim 10^{5} \,$\e, the detector is not perfectly linear even for well depths $< 10^{5} \,$\e.
Figure~\ref{saturation} (b) shows the correction factor for the linearity as a function of the integrated signal normalized at 100 \e.
Systematic separation lying low from the linearity might be caused by another reset anomaly, but this effect is small enough to neglect 
because the 1\% difference in the first several points will be reduced by a factor of 100 by fitting to the whole data set.
As described in section \ref{sS:calibration}, since the LRS absolute calibration was done with an integrated signal of $\sim 4\times 10^4 \,$\e
and the LRS diffuse integrated sky brightness is $\sim$ 2000 \e, 1.5\% linearity correction is required.

\subsubsection{Image Persistence}
\label{ssS:memory}

\begin{figure*}
\subfigure{
	 \includegraphics[scale=0.41, angle=90]{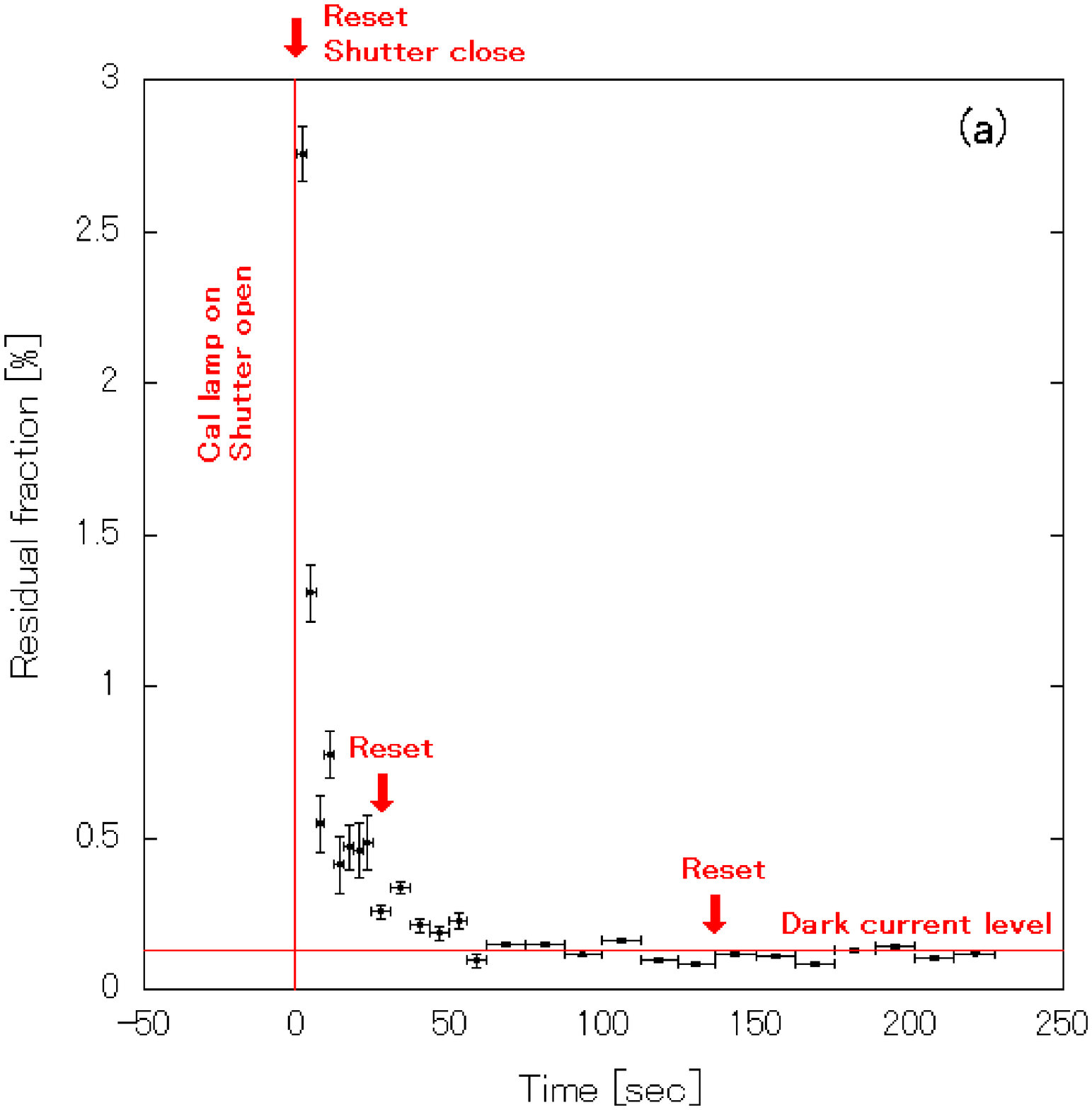}}
\subfigure{
	 \includegraphics[scale=0.41, angle=0]{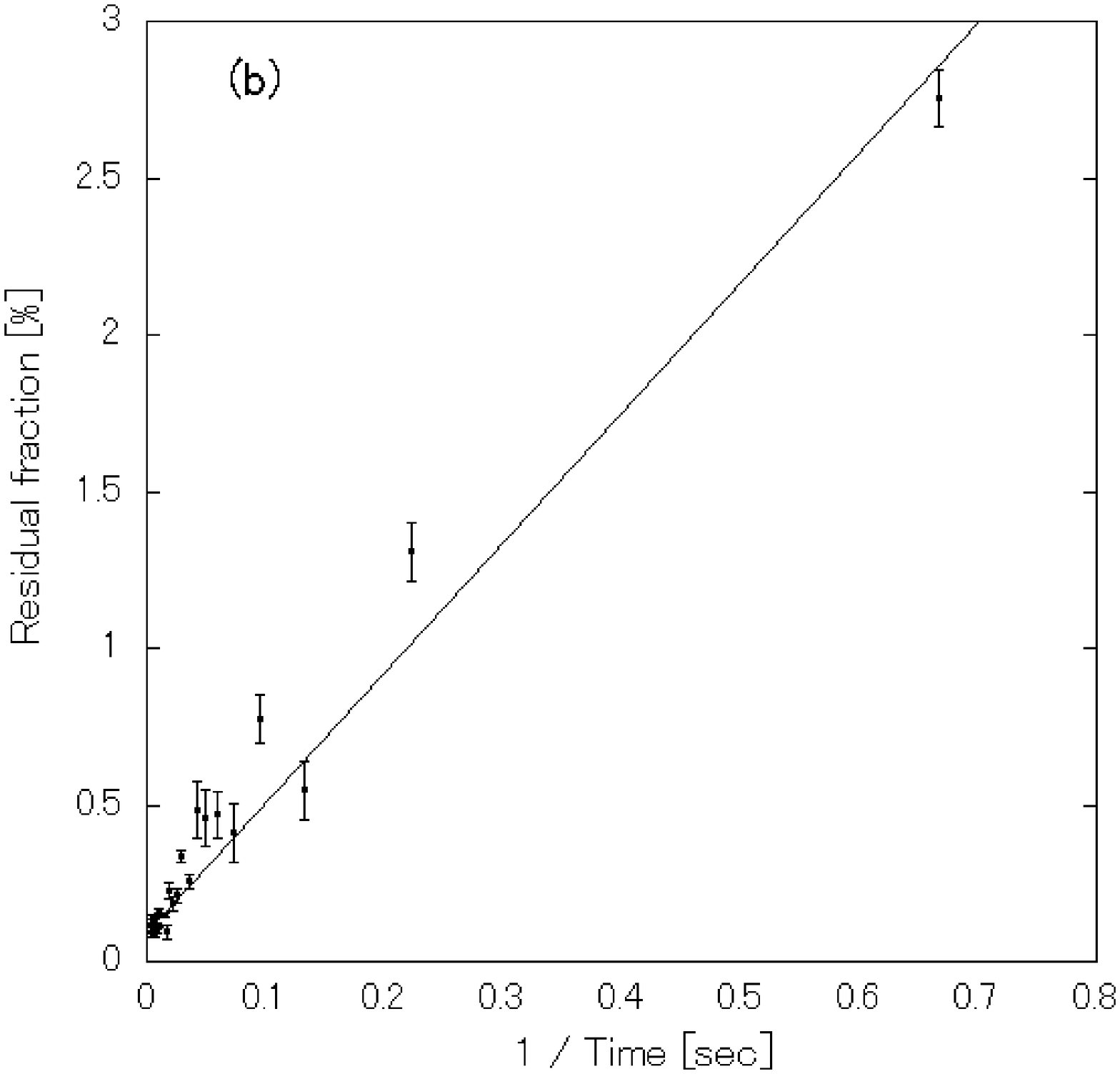}}
\caption{Behavior of the image persistence in the LRS detector. 
  Persistence signals at pixels where the average signal from the calibration lamp was 640 \eps\ are shown as a function of time (a) and the reciprocal of time (b). 
  The persistent signal is inversely proportional to time, and not reduced by multiple resets.
\label{memoryeffect}}
\vskip 0.4cm
\end{figure*}

PICNIC arrays can suffer from image persistence, a subtle increase in dark current in response to prior illumination.  
This image persistence varies significantly from detector to detector.
We studied this effect with the flight PICNIC detector.
The PICNIC detector was illuminated by the calibration lamp, and then the shutter was closed and dark images were taken. 
The first several dark frames have the image persistence from the calibration lamp,  
and the image persistence in the resulting dark images was studied as a function of time and the number of reset signals.
The dark images were made from 12, 25, and 50 raw frames which is equivalent to 3, 6, and $13 \,$s integrations, respectively. 

Figure~\ref{memoryeffect} shows the image persistence at pixels where the average signal from the calibration lamp was 640 \eps .  
As Figure~\ref{memoryeffect} shows, the image persistence after the first reset signal was $< 3$\% of the original value, 
and reduces to the dark current level inversely with time.
This behavior is similar in all pixels over the array.
Although there are widespread misconceptions that the image persistence can be also reduced by multiple resets,
we found multiple reset signals, except the first reset, were not effective in reducing persistence.
\citet{Smith08} introduced a qualitative model of the image persistence of the HgCdTe detector array.

Since the intervals between the fields during the CIBER flight was typically 15 seconds,
the image persistence is visible in flight data when very bright stars are detected, 
and the persistent signals were about 0.2\% of the original signals.
We can mask detected persistent stars in such cases.
\ \newline

\subsubsection{Electrical Cross-talk}
\label{ssS:crosstalk}

\begin{figure}
\begin{center}
\includegraphics[scale=0.9]{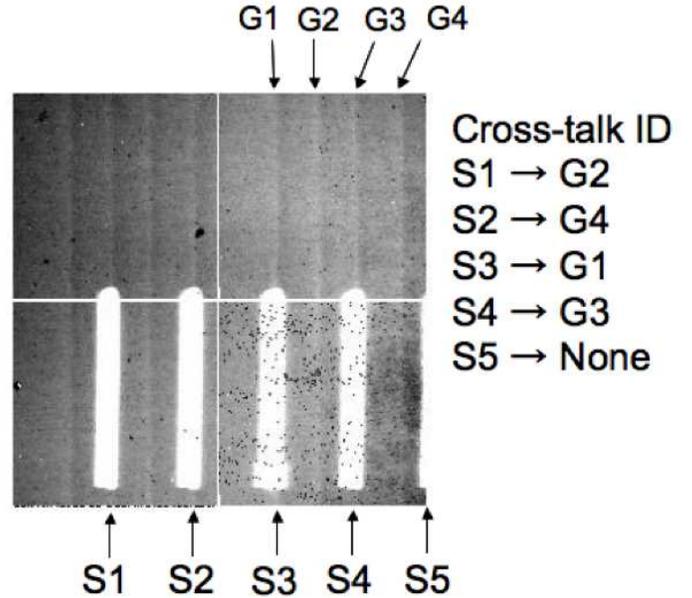}
\end{center}
\vskip -0.5cm
\caption{The image to ascertain the cross talk. The upper two quadrants of the detector array were masked and then the detector was illuminated by strong thermal emission from the laboratory.
         This figure clearly shows the cross-talk signals (labeled ``G'') at identical positions of the incident thermal signals (labeled ``S'').
  \label{crosstalk}}
\vskip 0.4cm
\end{figure}

HgCdTe detector array readouts typically suffer from some level of
electrical cross talk; indeed, we have observed cross-talk in the LRS electronic chain.  
A test was conducted to investigate the effects of the electrical cross-talk.
In this test, the upper two quadrants of the detector array were masked and then the detector was illuminated by strong thermal emission from the laboratory and data was acquired.
The cross-talk signals (labeled ``G'' in Figure \ref{crosstalk}) appeared even in the masked quadrants 
at identical positions of the incident thermal signals (labeled ``S'' in Figure \ref{crosstalk}).
This measurement showed that 0.35 \% of the incident signal on a quadrant is injected into the other quadrants.  
Although this fraction is significantly below the noise level under normal
flight conditions, the uncertainty in its correction imposed an
irreducible systematic error on the CNIRB measurement from the first
flight data set because of the large thermal stray light
(this is discussed in section \ref{ssS:baffle}).  

\subsection{Optical Evaluation}
\subsubsection{Focus Test and Imaging Quality}
\label{ssS:focus}

In order to maximize the number of independent pixels available for
the measurement of surface brightness, the detector array must be at
the best focus position of the instrument optics during flight.  
Because it is not possible to measure the positions of the optical
components in the LRS to sufficient accuracy to calculate the position
of the LRS's focal plane a priori, the position of the focal plane
must be determined experimentally.
To verify the focus of the instrument and measure the imaging quality at
the best focal position, we carried out a campaign of laboratory measurements. 
For these tests, a high-resolution monochromator system MS257 manufactured by
Newport\footnote{http://www.newport.com, model number 77000 and 70527}
is used as a light source with a narrow spectral band (see \citet{Zemcov11} for detail). 
The wavelength resolution of the monochromator is 15 nm in this test, 
so that spreading in the LRS dispersion direction is negligible.
The incoming beam from the source is passed through
a pinhole at the prime focus of a collimating telescope with a focal length
$f_{\mathrm{col}} = 910$ mm made by Parks Optical
Inc.\footnote{http://www.parksoptical.com}.  The focus of the
collimating telescope is determined by an auto-collimation procedure using a
collimating microscope and flat mirror placed at the aperture of the
collimator (see \citet{Zemcov11} for detail).  
Once in focus, the beam from the collimating telescope is
fed to the aperture of the LRS optics.

As the slit mask blocks most of the optical paths from the LRS
aperture to the detector array, the beam of the collimating telescope must be
steered to fall on the notch in the slit mask. 
The image formed on the array is a measurement of the PSF in response to a
point source.  As the detector is mechanically fixed in the
instrument, the best focus is determined by moving the pinhole along
the optical axis of the collimating telescope.  Given the focal
lengths of the LRS $f_{\mathrm{LRS}}$ and collimating telescope
$f_{\mathrm{col}}$, the relation between shifts along the optical axis
at the position of the pinhole $\Delta l_{\mathrm{pinhole}}$ from best focus
and the equivalent shift at the detector $\Delta l_{\mathrm{LRS}}$ can be
calculated by:
\begin{equation} 
\Delta l_{\mathrm{LRS}} = \left( \frac{f_{\mathrm{LRS}}}{f_{\mathrm{col}}} \right)^2 \Delta l_{\mathrm{pinhole}}  .
\end{equation}

\begin{figure}[h]
\begin{center}
\includegraphics[scale=0.4,angle=90]{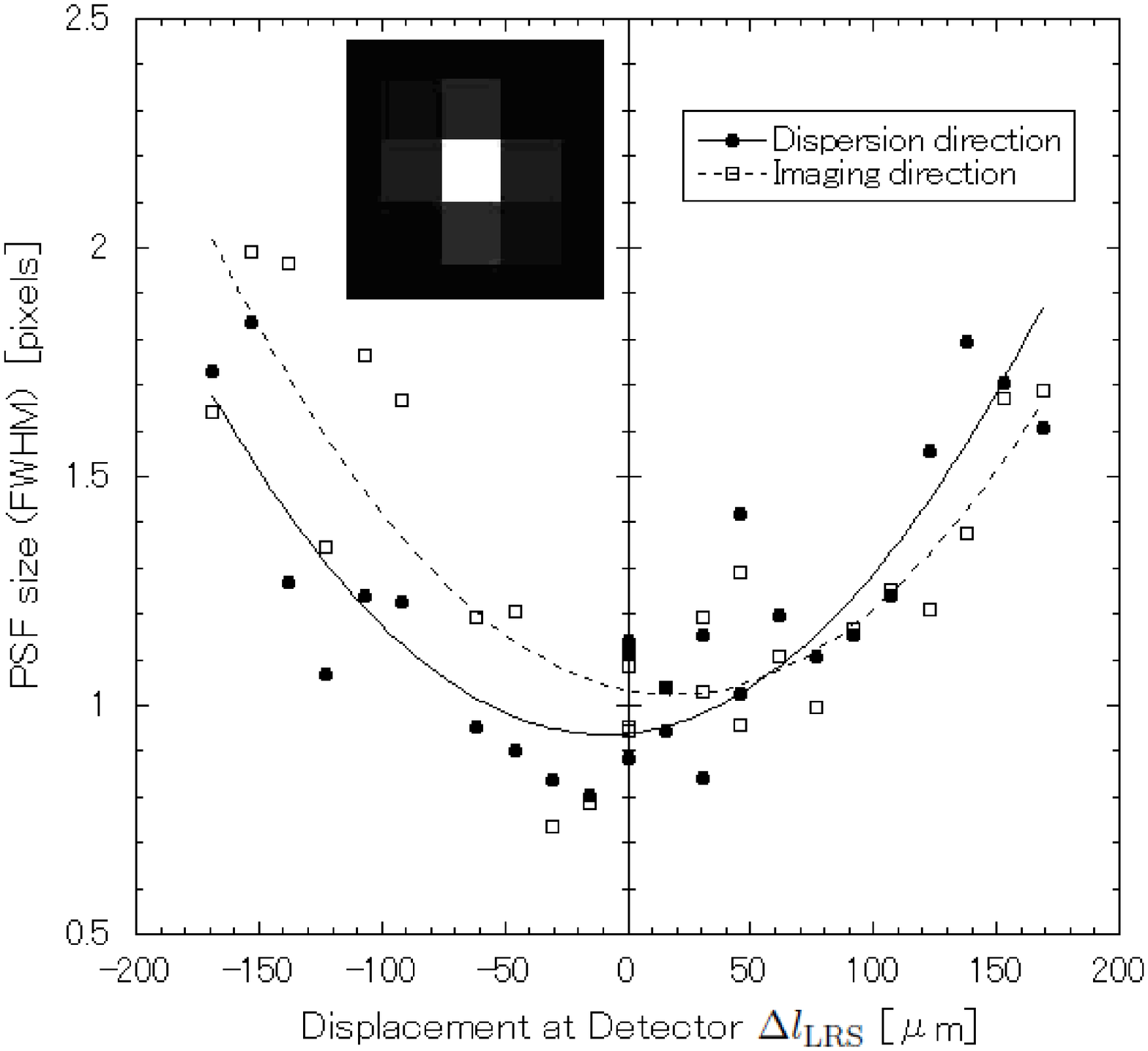}
\end{center}
\caption{Results of focus testing for the LRS configuration flown in
  CIBER's first flight.  Filled points indicate the size of the PSF in
  the direction of wavelength dispersion, and open points indicate the
  PSF size in the orthogonal imaging direction.  The solid and dashed lines
  show the best-fit quadratic curve for these populations,
  respectively.  The inset shows the resulting point source image; the
  best PSF size at this position is consistent with one pixel and the
  encircled energy in the main pixel is $80$ \%.
  \label{focus}}
  \vskip 0.4cm
\end{figure}

As the focal plane assembly is
statically mounted on the instrument, moving the detector array to the
best focus of the instrument requires opening the experiment and
physically shimming the focal plane assembly by $\Delta
l_{\mathrm{LRS}}$.  The experiment is then cooled and the test
performed again.  Due to the vagaries of the mechanical contraction
of the various components under cryogenic cycling, the position of the
detector array may not fall at the intended position.  The focus
position is therefore determined iteratively by testing, warming the
experiment and shimming the focal plane assembly, cooling, and testing again.
After several such iterations, the detector position repeatably falls
at the best focus of the instrument.  Figure~\ref{focus} shows the
best focus performance of the LRS in the configuration flown
during the first CIBER flight.   
The misalignment between spatial and dispersion directions in Figure~\ref{focus} was caused by undersampling.
Measurements of a pixel-scale spot size suffered from a large systematic error, depending on where the centroid of the stop is locating in a pixel.
However, it clearly shows that the detector was positioned close to the optimal focus where the PSF size was $\sim$1 pixel, 
showing that the imaging quality of LRS at the flight focus position matches the design. 
The focal depth of LRS is $\sim$80 $\mu$m (pixel size: 40 $\mu$m at $f/2$).
We found no evidence that any coma, astigmatism, vignetting or other effects reduce the sensitivity.

Prior to launch we measured instrument focus, subjected the experiment to a 
3-axis random vibration test, and then repeated the instrument focus measurements.
We did not observe any measurable change in focus due to vibration.

\subsubsection{Wavelength Calibration}
\label{ssS:pixeltolambda}

\begin{figure*}
\begin{center}
\includegraphics[scale=0.62,angle=0]{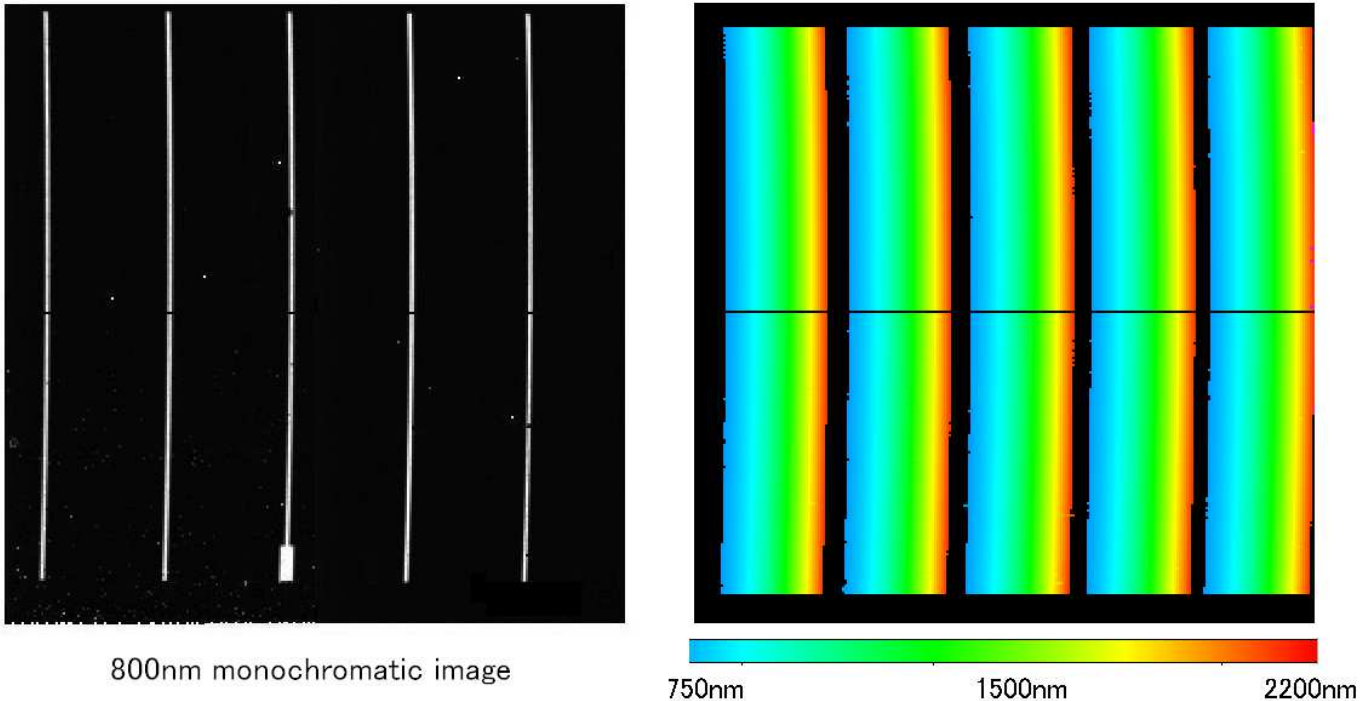}
\end{center}
\vskip -0.3cm
\caption{The left hand panel shows an image of the spectral test of
  800 nm; the slit image moves to the right as the wavelength increases.
  The right hand panel shows the wavelength map of the detector array
  obtained from this spectral test. \label{spectral}}
  \vskip 0.5cm
\end{figure*}

\begin{figure}
\begin{center}
\includegraphics[scale=0.37,angle=90]{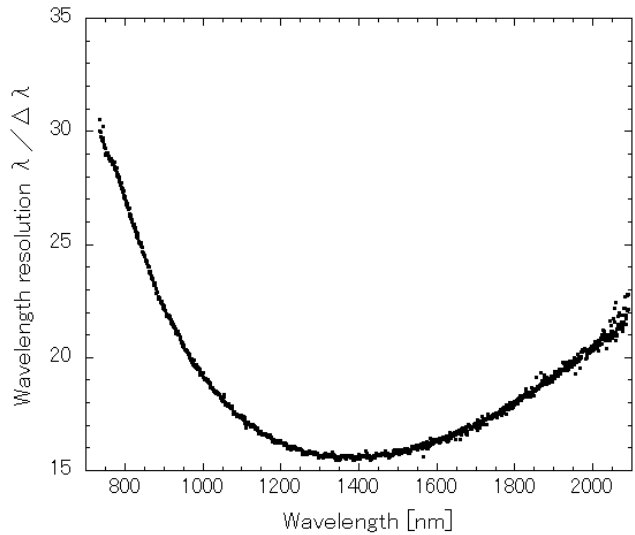}
\end{center}
 \vskip -0.3cm
\caption{The LRS wavelength resolution $\lambda / \Delta \lambda$ versus wavelength as measured from spectral characterization with a monochromatic light source.
             This wavelength resolution is almost same among the five slits. \label{resolution}}
\vskip 0.4cm
\end{figure}

In order to map pixels on the focal plane to effective wavelengths,
we measured the pixel-to-wavelength calibration using a spectral test.
The output of the monochromator discussed in Section \ref{ssS:focus}
is fiber-coupled to an integrating sphere, producing an
aperture-filling monochromatic light source (see \citet{Zemcov11} for detail about the integrating sphere).
The LRS views the aperture of the integrating sphere while the wavelength of the monochromator 
is stepped from 750 nm to 2100 nm in 5 nm increments.  
The wavelength resolution of the monochromator is 15 nm in this test,
which is smaller than the LRS resolution.
Figure~\ref{spectral} shows both an image obtained at a particular wavelength during this
spectral testing and the resulting wavelength calibration map derived
from an ensemble of such images over 0.75 $\mu$m $< \lambda <$ 2.1 $\mu$m.
Figure~\ref{resolution} shows the measured wavelength resolution of
LRS from this measurement, which is found to be consistent with the
design specifications.

As the slit mask provides five independent measurements of how light
propagates through the LRS optical chain, the optical performance
of the telescope and the tilt of the detector chip itself can be
monitored by comparing the width of the slit images.  These widths are
found to be constant over the detector array which indicates that the
design optical performance has been attained, and that no tilt
of the detector chip is evident. 

\subsubsection{Optical Baffling and Stray Light}
\label{ssS:baffle}

The limiting instrumental systematic in the first flight data was
thermal emission from the ambient rocket skin surfaces scattered into the telescope
aperture \citep{Tsumura10}.  The rocket skin was heated to
temperatures up to $400 \,$K by air friction during the powered ascent, 
and this thermal emission dominates over the astrophysical signal at wavelengths longer than 1.6 $\mu$m.  
Emission from the skin can enter the LRS optics by scattering on the LRS baffle and first lens. 
To reduce the sensitivity of the LRS to this stray light during CIBER's second flight,
the LRS baffle was redesigned and fabricated to be blacker in the NIR, 
and a pop-up baffle was added to extend past the skin and rocket door,
eliminating all lines of sight from the skin to either the interior of the LRS baffle or the first lens. 
A full discussion of these modifications can be found in \citet{Zemcov11}.
The performance of the new LRS baffle system was evaluated by an off-axis test, following the methodology of \citet{Bock95}.

\begin{figure*}
\subfigure{
	 \includegraphics[scale=0.4, angle=90]{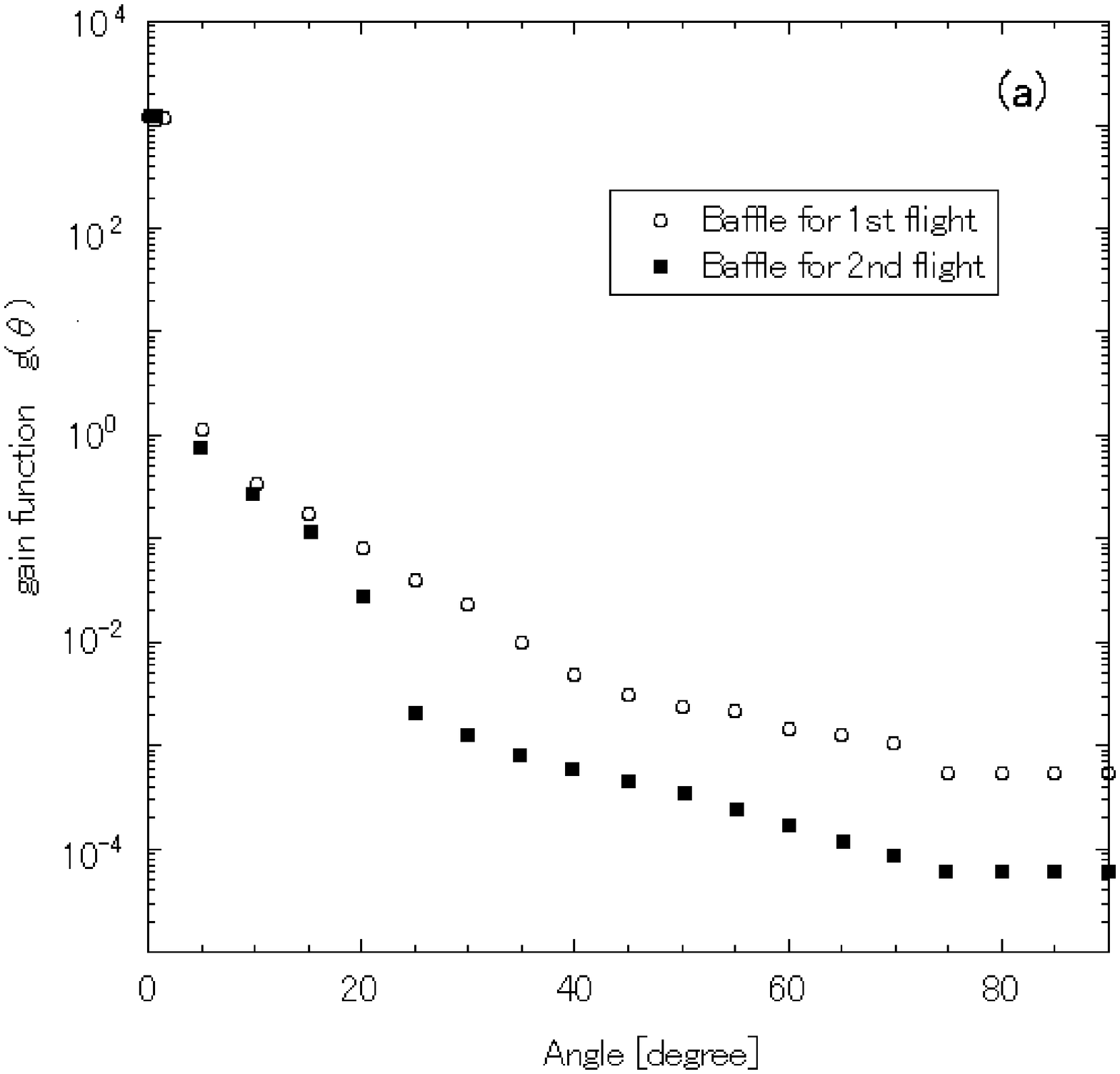}}
\subfigure{
	 \includegraphics[scale=0.4, angle=90]{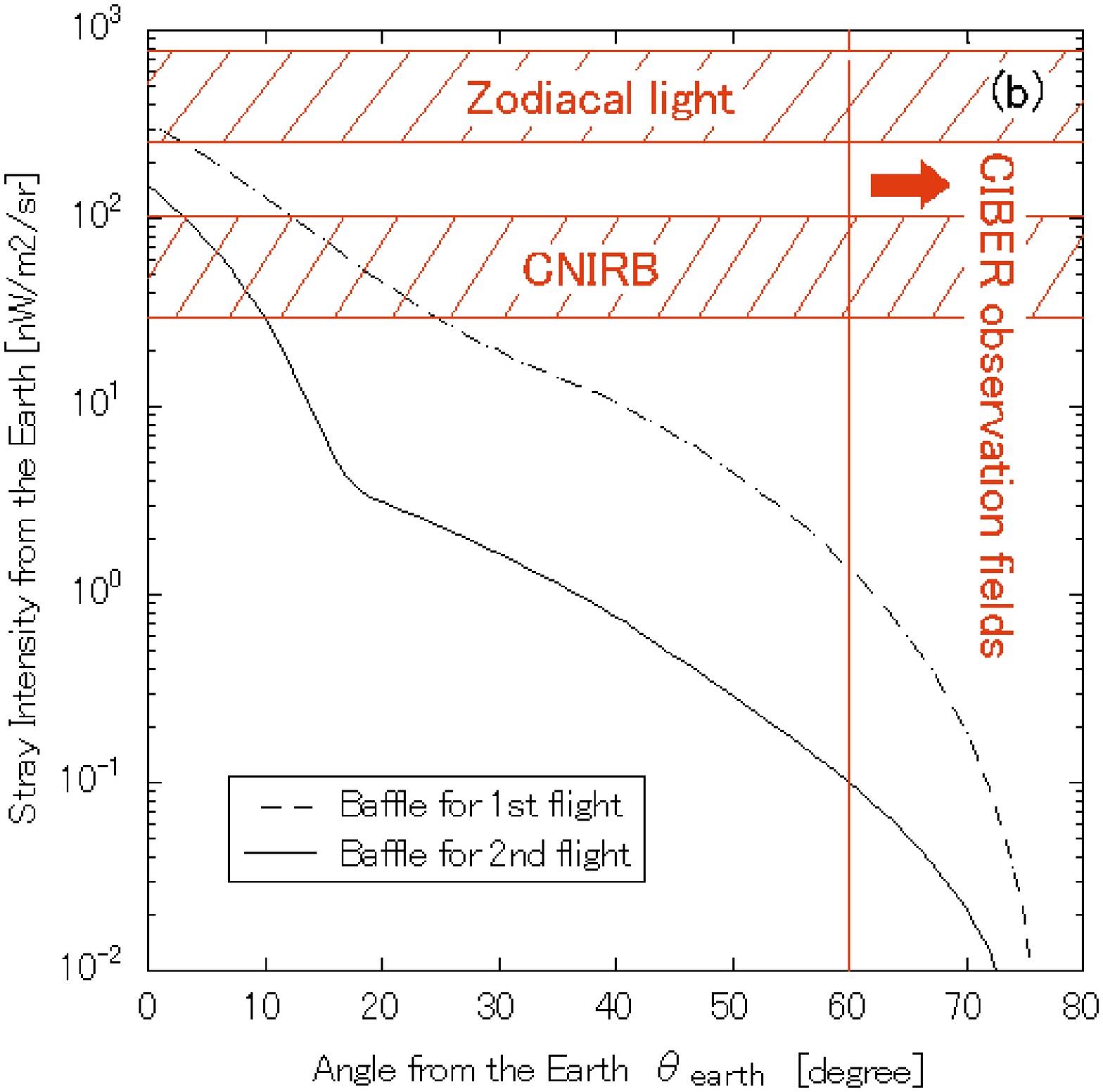}}
\caption{(a) The off axis response of the first and second CIBER flight
  LRS configurations.  The results for the baffling scheme flown in
  CIBER's first flight are shown in open circles, and the results of
  the same measurement for the modified baffling flown during the
  second flight are shown as filled squares.   
  (b) The calculated stray intensity from the earth as a function of the angle from the Earth
  with the old baffling scheme (dot-dashed line) and the modified baffling scheme (solid line).
  Brightness of ZL and CNIRB are also shown.
\label{baffle_fig}}
\end{figure*}

To implement this test, the LRS slit was removed 
and we replaced the PICNIC array with a silicon photo diode S10043 manufactured by Hamamatsu photonics.  
The active area of the photo diode is  $10 \,$mm$\times 10 \,$mm,
almost same size as the PICNIC active area.
The instrument was installed on a custom optical bench which can be
tilted from 0 to 90 degrees from horizontal.  Light from a halogen
lamp is chopped at $\sim 18 \,$Hz and coupled to the collimating telescope,
and illuminates the LRS entrance aperture and baffle.
The brightness of the source is
measured on-axis and then as a function of the angle between the incoming collimated
light and the LRS telescope, yielding a measurement of
the off axis light rejection of the new LRS baffling.  
Figure~\ref{baffle_fig} (left) shows the result of this baffle measurement in terms of the the gain function defined as
\begin{equation}
g(\theta) = \frac{4\pi }{\Omega }G(\theta)  
\end{equation}
where
$G(\theta)$ is the response to a point source from an off-axis angle $\theta$ normalized to unity on axis,
and $\Omega $ is the FoV solid angle in the measurement.
The gain function is independent of FoV and is thus useful for comparing optical systems designed to observe extended emission.
The modified baffling scheme provided an one order improvement for angles $>20^{\circ}$ due to the improved blacking and the new pop-up baffle.

\begin{figure*}
\begin{center}
	 \includegraphics[scale=0.3, angle=90]{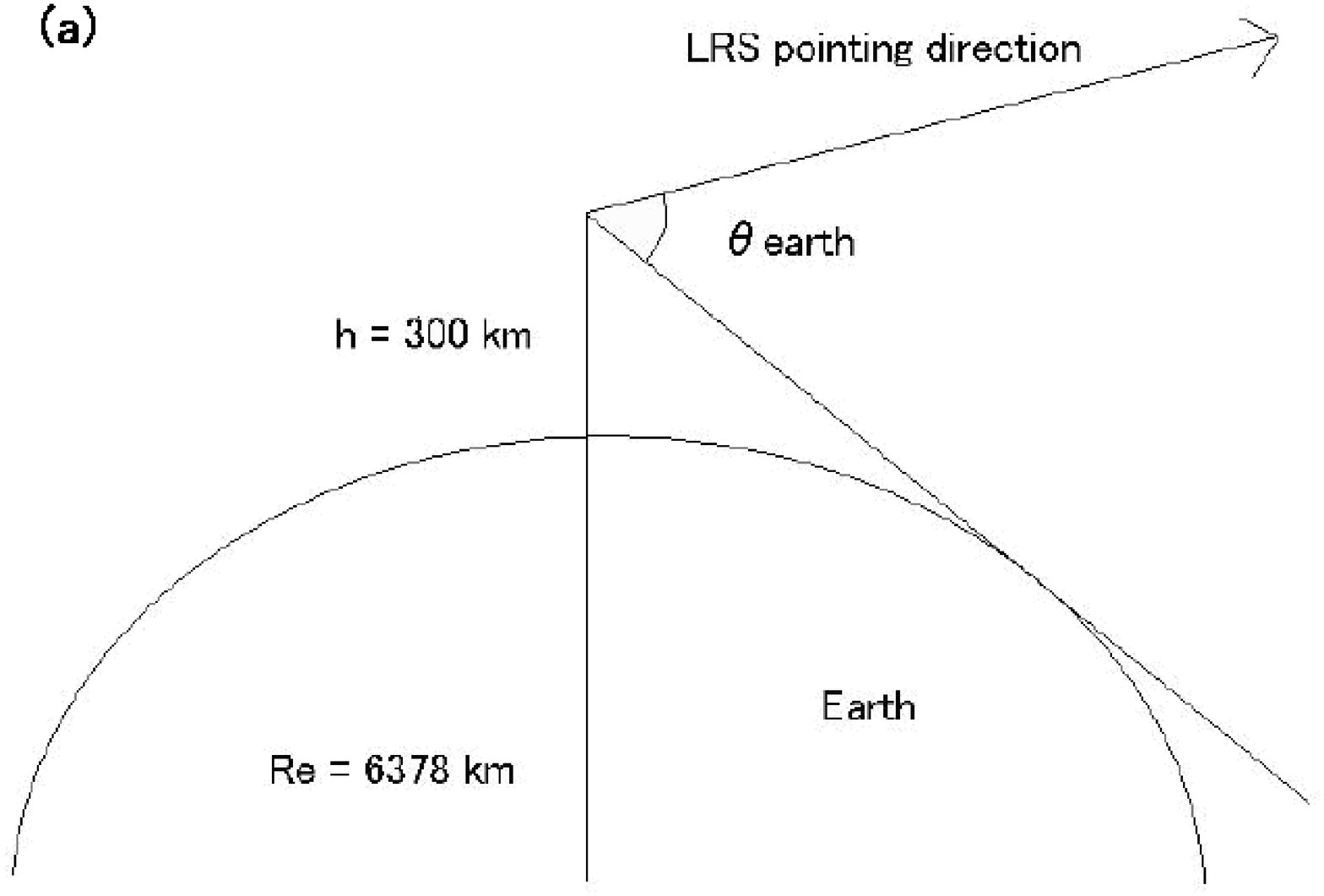}
	 \includegraphics[scale=0.3, angle=90]{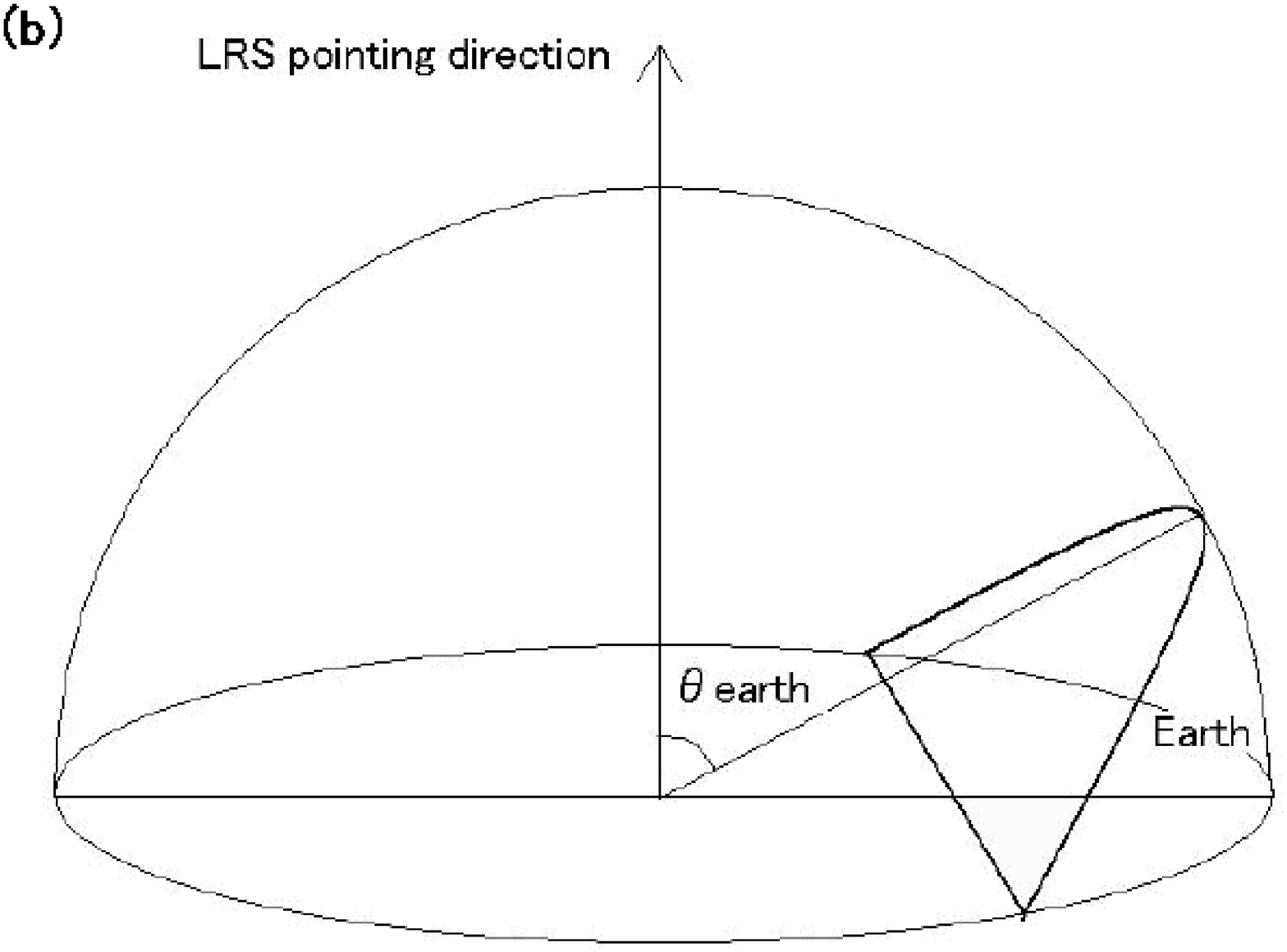}
\end{center}
\caption{Geometry model. Definition of angles between the Earth and CIBER (a) and the
         2$\pi$ steradian hemisphere used in the integral in equation (\ref{baffle_eq}) (b).
\label{earth}}
\vskip 0.4cm
\end{figure*}

Since the new pop-up baffle goes past the the end of the skin section, it blocks all stray paths from the rocket skin.
In addition, it also reduces stray light from large angle, especially from the Earth.
Here we estimate the stray light from the Earth with the old and new baffle scheme.
The apparent surface brightness from the earth referred to the sky $I_{stray}(\theta_{earth})$ is calculated by
\begin{equation}
I_{stray}(\theta_{earth})=\frac{1}{4\pi}\int g(\theta)I_{earth}(\theta,\phi)d\Omega  ,
\label{baffle_eq}
\end{equation}
where $\theta_{earth}$ is the angle between the LRS pointing direction and the Earth, and $I_{earth}(\theta,\phi)$ is the intensity from the earth.
For calculating the equation (\ref{baffle_eq}), we assumed a simple geometry model (Figure \ref{earth}) and $I_{earth}$ as
\begin{eqnarray}
I_{earth}(\theta,\phi)=\left\{ \begin{array}{ll}
6\times 10^4 \ \textrm{[\nw ]} \ \ \  $\citep{Leinert98}$\\
\ \ \ (\textrm{Airglow brightness from the Earth})  \\
0 \ \ \  (\textrm{out of the earth})  . \\
\end{array} \right.
\end{eqnarray} 
Note that the thermal radiation from the Earth (300K black body) is less than half of the airglow emission at 2.2 $\mu$m.
The equation (\ref{baffle_eq}) was calculated numerically and the result is shown in Figure~\ref{baffle_fig} (right) as a function of the angle from the Earth.
Since the minimum elevation angle in CIBER observation targets is $>$ 60 degrees, the estimated stray intensity from the Earth is negligible.

No other internal ghosts or reflections were found even in the case that the brightest star (42Dra) was in the FoV.

\subsection{Absolute Calibration}
\label{sS:calibration}

\subsubsection{Test configuration}
\label{ssS:configuration}

Since the primary scientific motivation for the LRS measurement is to
determine the absolute spectrum of the CNIRB, absolute calibration is
an essential component of the LRS instrument characterization.  An
important point for LRS absolute calibration is that a calibration of
sensitivity to extended emission is required.  Unfortunately, an accurate measurement of the
fluxes of individual sources in flight is made difficult by the
slit mask, which requires precise knowledge of the PSF and pointing for use in calibration.  
Therefore, we need to measure a calibrated
extended source in the laboratory for the LRS calibration.

To this end, we use instruments provided by the National Institute of
Standards and Technology (NIST) in a collaborative arrangement.
Two calibration measurements are performed using different light sources: 
a tunable laser (SIRCUS; \citet{Brown06}) and a quartz-tungsten-halogen lamp. 
The traveling SIRCUS lasers brought to calibrate CIBER consist of a Ti:sapphire laser tunable from 700 nm to 1000 nm, 
and an optical parametric oscillator, fed by the Ti:sapphire laser, which was used to cover the range from 1000 nm to 2100 nm.  
A quartz-tungsten-halogen lamp manufactured by Schott\footnote{http://www.us.schott.com/english/index.html, part number DCRIII} 
is used for the white light calibrations. 
A comparison of the monochromatic and white light measurements is an independent check 
that the calibration is correct and that there are no low-level broad-band light leaks. 
In either case, the light source is coupled to a 48-inch barium sulfate integrating sphere, the output port of which is viewed by the LRS. 
The transfer standard detectors are different for the two sets of measurements. 
NIST broad-band radiance meters calibrated at SIRCUS were used for the laser measurements, 
and the NIST Remote Sensing Laboratory's Analytical Spectral Devices (ASD) FieldSpec3 spectroradiometer was used for the white light measurements. 
These measurements are used to accurately calibrate both the absolute and relative (i.e. pixel-to-pixel) responsivity of the LRS.
A more detailed explanation of the CIBER calibration apparatus can be found in \citet{Zemcov11}.

The large mismatch in sensitivity of the LRS instrumentation compared to the NIST radiometers 
requires a scheme to attenuate the calibrated surface brightness to a level that does not saturate the LRS. 
The radiance levels observed by the LRS must be $10^4$ to $10^6$ fainter than that needed to make accurate measurements with the reference radiometers.  
The intensity levels for the LRS were reduced by passing light from the source through a cascade of one or more smaller integrating spheres 
(2 inches to 6 inches in diameter) in series with the large sphere, 
and using a monitor on one of the cascaded spheres to precisely measure each attenuation factor.  
A series of separate measurements at higher intensity levels were used to establish the ratio between the monitor and the radiance of the large sphere. 
This involved measuring the ratios of the radiance or monitor signal between each sphere and the next in the cascade. 
The flux levels were adjusted, as needed for adequate signal-to-noise ratios, over a very large dynamic range either by bypassing an integrating sphere in the cascade, 
adjusting the coupling between the light source and the optical fiber feeding an integrating sphere, or using a combination of half-wave-plate and a beam splitter for laser measurements. 

The broad-band radiometers used to make laser measurements were calibrated at NIST using a laser-fed integrating sphere source and a detector calibrated to measure optical power. 
The radiance levels of the source were deduced given a knowledge of the dimensions of apertures on both the source and detector and the distance between them. 
The responsivity of the radiometers varies smoothly as a function of wavelength allowing us to interpolate to the higher resolutions used during the CIBER measurements. 
With the exception of one radiometer, which has a lens, all were simple Gershun-tube designs, and all utilized single-element solid state detectors made of either InGaAs, Si or Ge. 
The ASD used for the white-light measurements was calibrated at NIST using an integrating-sphere source of known radiance that is traceable to fixed-point blackbodies.  

At each laser wavelength during the LRS measurements, the monitor signal was recorded with the laser on and then off (shuttered).  
The shutter intercepts the beam before it enters the fiber in order to measure the dark signals, 
and it is opened and closed by a trigger signal synchronized with the LRS integration. 
A similar procedure was used when the LRS viewed the white light source. 
Resets and periodic background measurements, made by blocking the lamp or turning it off, were performed manually.

\subsubsection {Data analysis}
\label{ssS:data analysis}

\begin{figure*}
\subfigure{
	 \includegraphics[scale=0.4, angle=0]{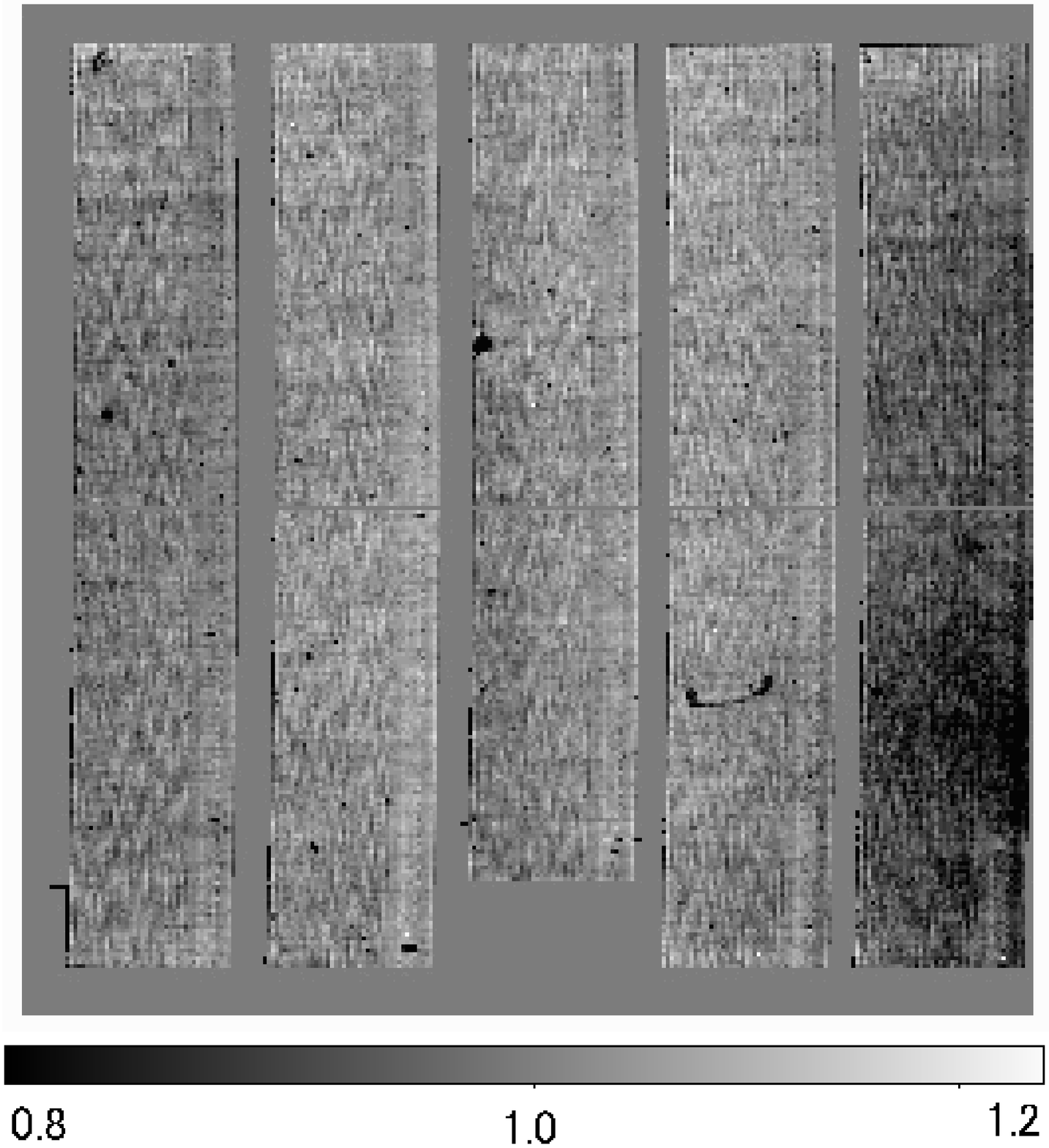}}
\subfigure{
	 \includegraphics[scale=0.4, angle=90]{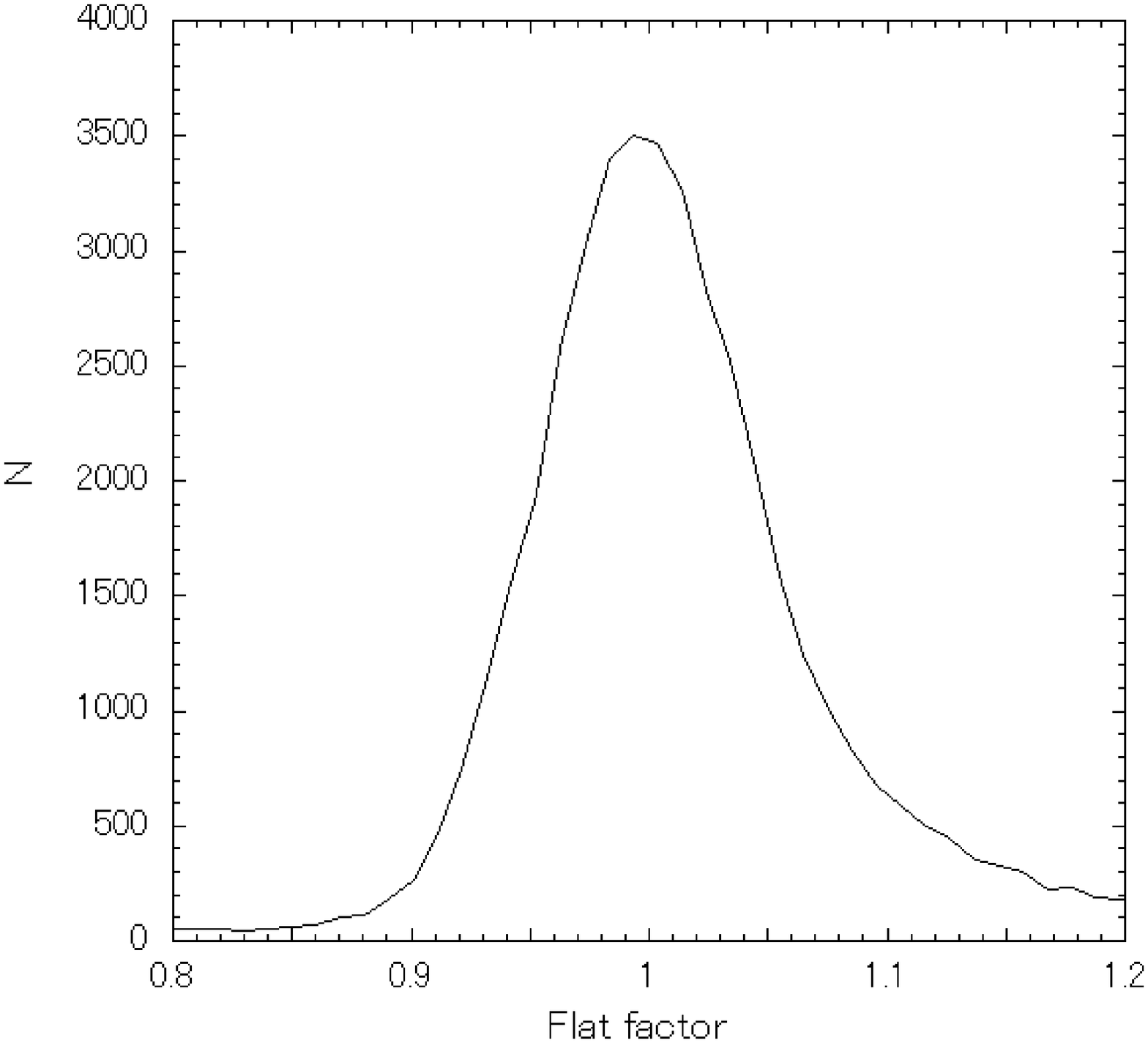}}
\caption{The pixel-to-pixel sensitivity correction map for the LRS (left) and its histogram (right).
The dispersion of the pixel sensitivity over the array is $\pm$5\% FWHM, 
and the quadrant-4 (lower-right) is less sensitive than the other quadrants.  
\label{flat}}
\vskip 0.5cm
\end{figure*}

\begin{figure}
\begin{center}
\includegraphics[scale=0.4,angle=0]{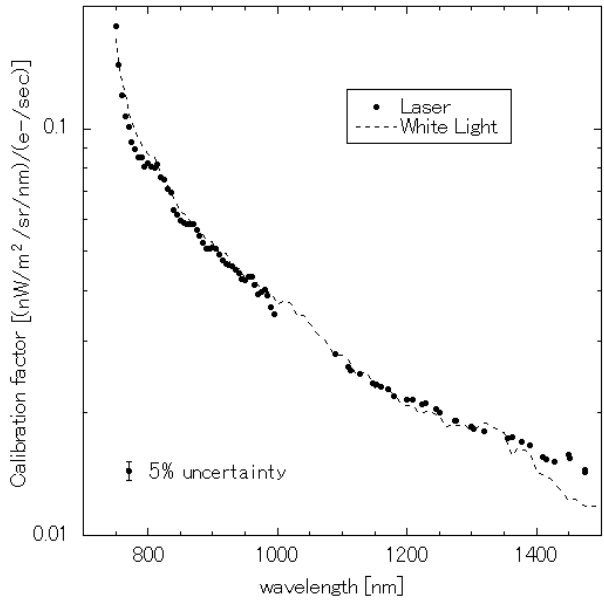}
\end{center}
\vskip -0.3cm
\caption{The LRS calibration curve.  Data points indicate the responsivity
  derived from the laser measurement, and the dotted line shows the
  responsivity derived from the broad-band measurement. 
  The error bar shows a $\pm 5 \,$ \% uncertainty. 
  \label{responsivity}}
  \vskip 0.4cm
\end{figure}

\begin{figure}
\begin{center}
\includegraphics[scale=0.4,angle=0]{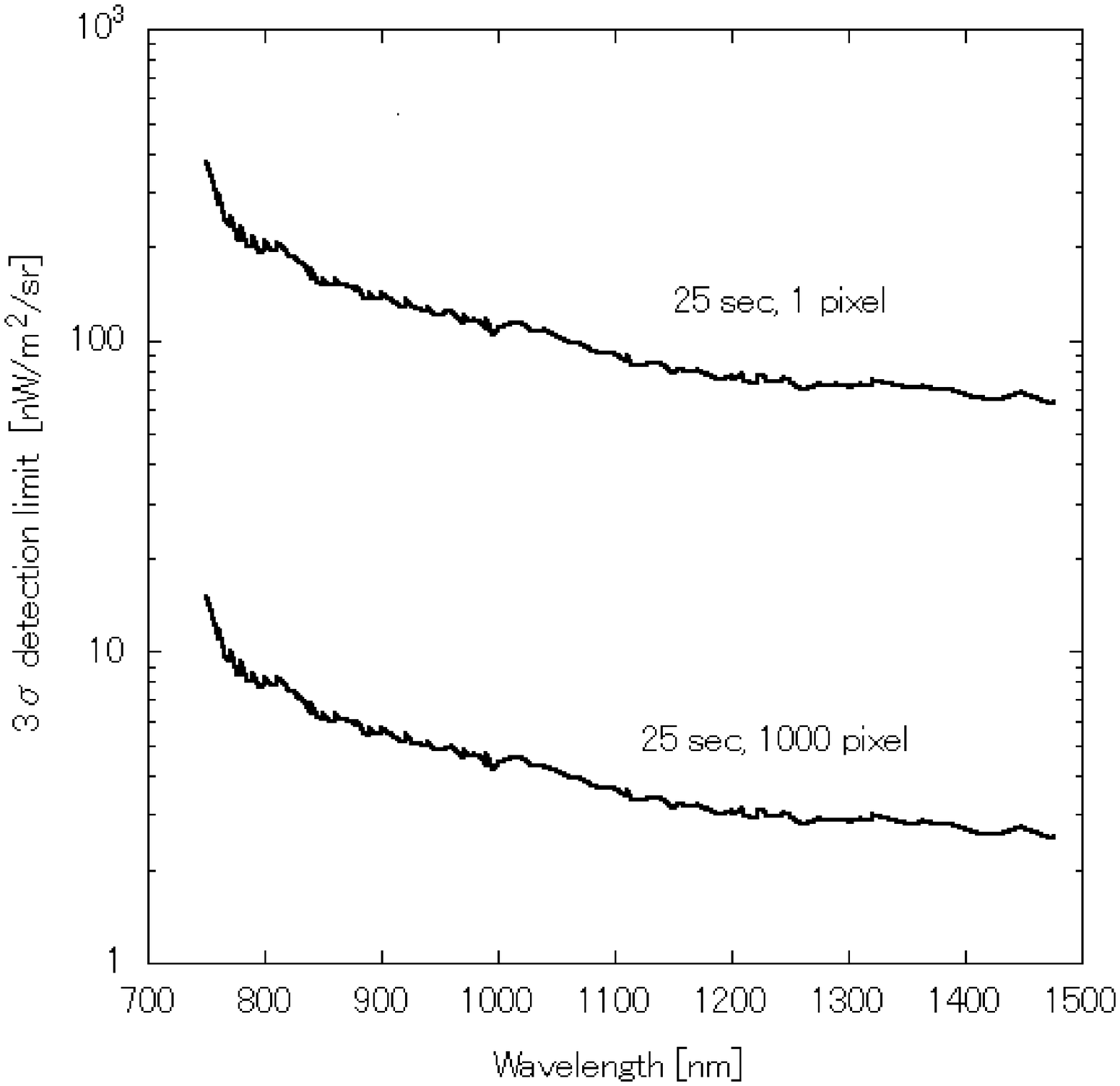}
\end{center}
\vskip -0.3cm
\caption{
  $3\sigma $ detection limit of the LRS for 25 second estimated from the noise level (read out noise + photon noise) in the flight data.  The sensitivity
  reaches  $< 10$ \nw\ which meets the requirement of LRS.\label{limit}}
  \vskip 0.4cm
\end{figure}

The calibration from the white light measurements is straightforward.
The ASD provides the absolutely calibrated spectrum $B(\lambda )$ of the integrating
sphere in \nw /nm; the LRS simply observes the source simultaneously
and obtains the equivalent signal $I_{\mathrm{LRS}}$ in \eps.  
The conversion factor $CF(x,y)$ from \eps\ to \nw /nm at pixel position $(x,y)$ can be calculated by computing the ratio of
these data;  
\begin{equation} 
CF(x,y) = \frac{T(\lambda )B(\lambda )}{I_{\mathrm{LRS}}(x,y)}
\end{equation}
where $T(\lambda )$ is  the transmittance spectrum of the test window (a BK7 parallel plate with 20mm thickness) of the CIBER cryostat which was measured separately,
showing $T(\lambda ) \sim 0.9$ across the LRS free spectral range.
The pixel position $(x,y)$ is related to wavelength $\lambda $ by the wavelength map 
shown in the right hand panel of Figure~\ref{spectral}.  
Because the wavelength resolution of the
LRS and reference ASD detector are different, the ASD measurements are
smoothed to match the wavelength resolution of the LRS.  These
smoothed spectra yield the white light calibration factor for the LRS.
A map of the pixel sensitivity, normalized to the mean over the array at each wavelength, is shown in Figure~\ref{flat}. 

The LRS calibration from the SIRCUS source is performed as follows.
Although light from the laser is essentially monochromatic, the
signal measured at the LRS detector is extended in the dispersive
($x$) direction due to the finite slit width ($\sim 2$ pixels) of the LRS.
To collect all of the dispersed power, the LRS signal is summed along the dispersion
direction on each slit and then compared to the monochromatic laser radiance $P(\lambda )$ in \nw.
\begin{equation} 
CF(y, \mathrm{slit}) = \frac{T(\lambda )P(\lambda )}{\bigtriangleup \lambda _{\mathrm{LRS}} \sum\limits_x I_{\mathrm{LRS}}(x,y)}
\end{equation}
To directly compare the laser result with the white light result,
the laser radiance was divided by the wavelength resolution of LRS $\bigtriangleup \lambda _{\mathrm{LRS}}$
(see Figure~\ref{resolution}) for matching units in \nw /nm.

We conducted this measurement three times, in 2008, 
2009 (for the 1st flight), and 2010 (for the 2nd flight). 
Between these measurements, the instrument was partially disassembled for servicing,
including a change of the prism to remove a stray light path.
The results of the laser measurement were consistent with each other within 5 \%
although some variation must be expected from servicing the instruments. 
However, the results of the white light measurement varied up to 30 \% in the worst case
due to bad repeatability of the absolute value from the white light measurements. 
Therefore, we rely on the result from the laser measurement 
and the result from the white light measurement was used for interpolating between laser bands.
The 5 \% calibration variation could be due to the instrument disassembly and modifications between two measurements.
Data analysis at $>$ 1.5 $\mu$m is still in process because the calibration of the long-wavelength detectors is still being measured at NIST.  
The average conversion curve over the array is shown in Figure~\ref{responsivity}.
Since this conversion curve was derived from the measurements  with an integrated signal of $\sim 4\times 10^4$ \e,
a linearity correction up to 1.5 \% is applied to compare with the flight data with the integrated signal of $\sim$2000 \e\ from the sky.

\section{Summary}
\label{S:summary}

The LRS is one of three CIBER instruments.  
The primary scientific motivation for the LRS measurement 
is to determine the absolute spectrum of the diffuse sky brightness.
The LRS is a 5-cm refracting telescope with a prism, operating at 0.75 $\mu$m $< \lambda <$ 2.1 $\mu$m with a spectral resolution of $\lambda /\Delta \lambda \sim$15-30. 
We evaluated the performance of LRS in the laboratory as described in this paper and found the following:

\begin{enumerate}
 \setlength{\itemsep}{-1pt}
 \item The noise performance was consistent with design predictions.
 \item A linearity correction up to 1.5\% is applied to compare the flight data with the laboratory calibration.
 \item Image persistence was confirmed and is detectable following observations of bright stars at the level of $\sim$ 0.2 \%.
 \item A small electrical cross-talk of the incident signal injects $\sim$ 0.35 \% into other quadrants.
 \item The best fit PSF size is consistent with one pixel and the encircled energy in the main pixel was $>$80\%.
 \item The modified baffling scheme for stray thermal emission removal provided a dramatic improvement for angles $> 20^{\circ}$, and the stray light signal from the Earth is negligible for the range of angles observed in flight.
 \item The calibration uncertainty is approximately $\pm$5\%.  
\end{enumerate}

Combining all instrumental uncertainty described in this paper and the read noise obtained in flight gives a demonstrated sensitivity of $< 10$ \nw\ for a 25 second integration, 3$\sigma $, 1000 pixels as shown in Figure~\ref{limit}.
The dominant component of uncertainty, $\pm$5\%, is from the instrument calibration.
In the actual astronomical data analysis, there are other astronomical error elements rather than the instrumental systematics. 
The dominant error elements from astronomy come from the estimation of the ZL and DGL brightness to subtract.
Details of such astronomical error evaluation will be discussed in future science papers.

The first result of LRS from CIBER 1st flight was summarized and published in \citet{Tsumura10} based on the characterizations described in this paper,
although some of them have been modified from the first flight to improve its performance.
The 2nd flight data is under analysis now and the result will be reported in forthcoming papers.
Owing to the modifications especially to combat the thermal stray light, data with good quality enough to obtain the CNIRB spectrum was acquired in the 2nd flight.

\section*{acknowledgments}

This work was supported by NASA APRA research grants NNX07AI54G, NNG05WC18G,
NNX07AG43G, and NNX07AJ24G.  Initial support was provided by an award to J.B.
from the Jet Propulsion Laboratory's Director's Research and Development Fund.
Japanese participation in CIBER was supported by KAKENHI (20$\cdot$34, 18204018,
19540250, 21340047 and 21111004) from Japan Society for the Promotion of Science
(JSPS) and the Ministry of Education, Culture, Sports, Science and Technology (MEXT).
Korean participation in CIBER was supported by the Pioneer Project from Korea
Astronomy and Space science Institute (KASI).

We would like to acknowledge the dedicated efforts of the sounding rocket staff at the
NASA Wallops Flight Facility and the White Sands Missile Range.  We also acknowledge the
work of the Genesia Corporation for technical support of the CIBER optics.  
K.T. acknowledges support from the JSPS Research Fellowship for the Young Scientists,
M.Z. acknowledges support from a NASA Postdoctoral Fellowship, and A.C. acknowledges
support from an NSF CAREER award.
We thank the referee for their useful suggestions which have improved this manuscript.

Certain commercial equipment, instruments, or materials are identified in this paper to foster understanding. 
Such identification does not imply recommendation or endorsement by the National Institute of Standards and Technology, 
nor does it imply that the materials or equipment identified are necessarily the best available for the purpose.

\end{document}